\documentclass[11pt]{article}

\usepackage[margin=1in]{geometry}
\usepackage[utf8]{inputenc}
\usepackage[T1]{fontenc}
\usepackage{lmodern}
\usepackage{textcomp}
\usepackage{microtype}
\usepackage{enumitem}

\usepackage{booktabs}
\usepackage{tabularx}
\usepackage{array}
\usepackage{amsmath}
\usepackage{graphicx}
\usepackage{xurl}
\usepackage[
    colorlinks=true,
    breaklinks=true
]{hyperref}

\usepackage{longtable}
\usepackage{caption}

\usepackage[
    backend=biber,
    style=authoryear,
    sorting=nyt,
    maxcitenames=2,
    mincitenames=1,
    maxbibnames=99,
    uniquename=false,
    uniquelist=false
]{biblatex}

\counterwithout{table}{section}

\begin{document}

\noindent
{\Large\bfseries
Large Eddy Simulation of Plunging Flows in Laboratory-Scale Bedrock Rivers
\par}

\vspace{0.5cm}

\noindent
Jayanga T. Samarasinghe\textsuperscript{1,2},
Laura V. Alvarez\textsuperscript{1,2},
Max Hurson\textsuperscript{3}, and
Jeremy G. Venditti\textsuperscript{3}

\vspace{0.4cm}

\noindent
\textsuperscript{1}Department of Earth, Environmental and Resource Sciences,
University of Texas at El Paso, El Paso, TX 79968, USA

\vspace{0.15cm}

\noindent
\textsuperscript{2}Environmental Science and Engineering Program,
University of Texas at El Paso, El Paso, TX 79968, USA

\vspace{0.15cm}

\noindent
\textsuperscript{3}School of Environmental Science,
Simon Fraser University, Vancouver, BC, Canada

\vspace{0.4cm}

\noindent
\textbf{Corresponding author:}
Jayanga T. Samarasinghe
(\href{mailto:jthambanged@miners.utep.edu}
{jthambanged@miners.utep.edu})

\vspace{0.6cm}

\noindent
\textbf{Key Points:}

\begin{itemize}[leftmargin=0.25in, itemsep=0.25em, topsep=0.25em]
    \item Large eddy simulations reproduce plunging flows in bedrock rivers
    and reveal three-dimensional turbulence structures at resolutions
    unattainable through field or laboratory methods.

    \item Channel constriction and discharge jointly control plunging-flow
    intensity, with moderate constrictions producing the strongest localized
    bed shear stress and erosion potential.

    \item Plunging flows are intermittent, with turbulence and bed stress
    peaking at the pool entrance and driving spatially concentrated bedrock
    erosion.
\end{itemize}

\vspace{0.4cm}

\newpage

\section*{Abstract}
Non-uniform flow dynamics in bedrock-bound channel morphologies play a critical role in landscape evolution because these reaches are locations along river long profiles where active bedrock incision occurs. Field observations indicate that plunging flows, characterized by velocity inversions within bedrock-bound constriction–pool–widening (CPW) channel morphologies, drive incision at the local scale. These flows generate high shear stresses that promote sediment transport and contribute to the development and maintenance of CPW morphology.  Previous studies of plunging flows have relied on coarse-scale field observations and labor-intensive laboratory experiments to investigate their dynamics. Here, we use eddy-resolving computational fluid dynamics models to examine plunging-flow behavior, building on experimental evidence that lateral channel constriction induces plunging flows. Using large-eddy simulations (LES) of laboratory-scale flows, we found that the optimal constriction for generating plunging flows is approximately 35\% under lower-flow conditions but increases to 50\% at higher flows because of changes in inlet velocity and flow depth. At higher discharge rates, channel constriction further amplifies the plunging effect, producing substantial shear stresses near the point of velocity inversion.  Increasing constriction also leads to greater velocity variance and more intermittent pulsing of plunging flows, both of which are likely to enhance incision potential. These findings highlight the need to refine bedrock incision models to better represent the dynamic and complex nature of plunging flows, moving beyond the simplified steady-flow assumptions that underpin most landscape evolution models.\\

\noindent\textbf{Keywords:} Bedrock Rivers, CFD, LES, Plunging Flow

\section*{Plain Language Summary}
Constriction–pool–widenings (CPWs) are bedrock-bound channel morphologies formed by flow structures known as plunging flows. Previous research has focused on identifying plunging flows and investigating the conditions that generate them, but studies of their dynamics have been limited by the scarcity of field observations, which are logistically challenging to obtain, and by labor-intensive laboratory experiments. Here, we use a computational fluid dynamics (CFD) model to simulate laboratory-scale plunging flows and examine their dynamics at high spatial and temporal resolution, which is difficult to achieve through field or laboratory observations alone. Our simulations show that laboratory-scale plunging flows develop when flow passes through a channel constriction, generating a jet-like flow that plunges toward the bed. The optimal constriction width for producing a scour pool is approximately 35\%, much lower than constrictions often observed in the field. This work opens the possibility of applying CFD models to field-scale bedrock rivers, enabling investigation of plunging-flow dynamics and erosion processes during high-flow events, where direct observations are often impossible to obtain.

\newpage
\section{Introduction}
Flow in bedrock rivers is often characterized by complex hydraulic phenomena, particularly when these rivers are bedrock-bound. Channel constrictions in bedrock canyons lead to the formation of deep pools and downstream widening, producing constriction-pool-widening (CPW) morphologies \parencite{hurson_abundance_2025,kusack_experiments_2024,venditti_struggles_2020,venditti_morphodynamics_2025,wright_covariation_2022}. When high-velocity water is funneled through a constricted section of a river, a high-velocity core develops through the constriction, which then follows along the riverbed, resulting in a velocity inversion where the faster-moving flow is at the bottom of the river and the slower-moving flow occurs at the surface \textcite{venditti_flow_2014}. This inversion generates a strong downward-directed flow along the channel centerline. To counterbalance this plunging motion, upwelling currents along the canyon walls bring slower, highly turbulent water from near the bed to the surface, forming large, intermittent flow structures, such as visible boils, that dissipate as the channel widens downstream \parencite{venditti_flow_2014}. \\

\noindent The mechanisms governing plunging flow formation remain poorly understood and existing hypotheses offer a range of interrelated explanations. \cite{venditti_flow_2014} proposed that plunging flows are driven by convective deceleration, wherein coherent three-dimensional flow structures transport slower-moving fluid toward the surface while pushing a core of maximum velocity (CMV) towards the bed. \cite{hunt_experiments_2018} attributed the formation of plunging flow to convective deceleration but emphasized additional controlling factors, including entrance bed topography, lateral constriction, pool entrance slope, discharge, and the width-to-depth ratio of the channel. Their laboratory experiments suggested that laboratory-scale flows transition to a trans-critical flow regime through a constriction, whereas field observations consistently indicate subcritical flow conditions \parencite{wright_covariation_2022}. \cite{cao_experiments_2022} argued that plunging flows arise due to backwater effects upstream of constrictions, forming a sediment deposit upstream of a constriction. They argued flow then spills over the deposit through the constriction, causing water and sediment to be drilled into the bed, scouring a downstream pool. \cite{hurson_amplification_2022} reinforced this explanation, demonstrating that at higher discharges, the increased flow through a constriction enhances the backwater and strengthens the plunging intensity. Despite these varying perspectives, no unifying framework fully explains the conditions under which plunging flows develop, highlighting a critical gap in our understanding.\\

\noindent Several phenomena in hydraulic engineering are similar to plunging flows in bedrock canyons, though direct analogues remain elusive. Engineered supercritical flows, such as those over broad-crested weirs cause flow to plunge below the water surface downstream of the weir, a behavior often referred to as a submerged hydraulic jump \parencite{fritz_hydraulics_1998, kabiri-samani_hydraulic_2010,leutheusser_backward_2001, wu_submerged_1996}. Additionally, the phenomenon of ‘choke flow’, where flow is intentionally super-elevated upstream of a constriction to become supercritical and subsequently plunge as an impinging jet, provides another related example \parencite{fritz_hydraulics_1998, wu_submerged_1996}. Notwithstanding these similarities, the key distinction lies in the nature of the flows through constriction-pool-widening observed by \cite{venditti_flow_2014} which are universally subcritical even at high flow conditions \parencite{hunt_experiments_2018,venditti_struggles_2020, wright_covariation_2022,hurson_amplification_2022,hurson_abundance_2025}. This subcritical nature contrasts with the supercritical flows associated with engineered structures. There are other natural flows that may be analogues to plunging flows including flow through pool-riffle sequences, where increased near-bed velocities in pools compared to riffles have been noted \parencite{keller_areal_1971}, however, the plunging flow phenomena described by \cite{venditti_flow_2014} have not been documented within such sequences. Thus, while engineered choke flows may provide valuable insights into flow in bedrock canyons (e.g. \cite{kieffer_1983_1985,kieffer_geologic_1989}),  the precise mechanisms driving natural plunging flows in bedrock rivers remain poorly understood and differ from those observed in engineered settings and other hydraulic phenomena.\\

\noindent Most observations of plunging flows come from large bedrock rivers, where logistical challenges constrain comprehensive field studies. The difficulties of accessing remote bedrock canyons and the financial costs associated with data collection often limit observations to one or two flow conditions. For instance, field observations have focused on low to moderate discharge levels, which do not fully capture the range of hydraulic behaviors and erosive processes occurring during higher flow when vertical bedrock incision may be most active \parencite{hurson_abundance_2025,hurson_amplification_2022, li_bed_2022, venditti_struggles_2020,venditti_flow_2014}. Additionally, the turbulent and dynamic nature of bedrock river flow complicates these studies, as extensive averaging and interpolation are needed to discern mean flow structures from field observations \parencite{venditti_flow_2014,hurson_amplification_2022,hurson_abundance_2025}. In contrast, laboratory experiments provide a controlled environment that allows for the exploration of a broader range of flow conditions and channel morphologies, but the inherent need for scale reduction can distort flow dynamics and are labor intensive. The fact similar flow phenomena are observed in the field and laboratory with different flow regimes highlight the limitations of scaled-down models \cite{hunt_experiments_2018}. Numerical modeling emerges as a powerful alternative for studying plunging flows, allowing for the simulation of a wide range of flow conditions and channel morphologies. These models facilitate the systematic variation of parameters such as discharge, sediment load, and bedrock geometry, enabling the exploration of complex three-dimensional flow structures without the logistical constraints of fieldwork or the scaling issues inherent in laboratory experiments. Moreover, numerical models provide valuable insights into long-term patterns and flow dynamics in bedrock rivers, offering explanations rooted in flow physics and physical laws that are difficult to obtain through traditional field observation.\\

\noindent Computational Fluid Dynamics (CFD) models serve as effective tools for simulating flow phenomena based on the fundamental principles of fluid mechanics. These models provide critical insights into the causality of various flow and transport processes, including mass and momentum transfer \parencite{xia_applications_2002}. In the context of fluvial geomorphology, understanding the resolved turbulence within these models is particularly important for predicting flow behavior and its influence on sediment transport \parencite{keylock_method_2011,keylock_theoretical_2005}. Several CFD modeling techniques are employed to simulate anisotropic turbulence at both field and laboratory scales, with the most common being eddy-resolving models based on Large Eddy Simulations (LES) and Detached Eddy Simulations (DES) \parencite{alvarez_detached_2017,alvarez_eddy-resolving_2021,espana_grid_2025,flora_uncertainty_2022, le_large-eddy_2019,zhang_significance_2021}. While Direct Numerical Simulations (DNS) are considered the most accurate eddy-resolving technique, they are computationally intensive and challenging to implement due to the complexity of defining accurate initial and boundary conditions, especially in studies involving high Reynolds number flows or complex geometries at field scales \parencite{chen_openfoam_2014,sibil_analysis_2024,yang_numerical_2024}. Consequently, DNS is frequently unsuitable for large-scale applications, despite its superior accuracy.\\

\noindent Given these constraints, LES emerges as a more feasible alternative for these types of studies. LES models, though less precise than DNS, still provide a reliable and efficient method for capturing the essential features of turbulent flows, particularly in the context of fluvial geomorphology \parencite{alvarez_detached_2017,alvarez_eddy-resolving_2021,dalessandro_large-eddy_2025,dashtpeyma_plunging_2023, flora_uncertainty_2022}. LES is well-suited to capturing the energy cascade and resolving turbulence above the sub-grid scale, both of which are central to understanding flow dynamics in bedrock rivers. LES therefore provides an effective compromise between accuracy and computational demands. Over approximately the past two decades, LES has emerged as a leading methodological framework for investigating turbulent patterns and eddy structures in field-scale river systems. While initially applied extensively in laboratory flume studies \parencite{bayon_performance_2016,dashtpeyma_plunging_2023,zhou_large-eddy_2017}, LES has since demonstrated significant adaptability and effectiveness in capturing complex three-dimensional flow physics at field scales \parencite{alvarez_turbulence_2015,alvarez_detached_2017, alvarez_eddy-resolving_2021,espana_grid_2025}. This combination of practicality and reliability positions LES as the optimal choice to study turbulent flow structures and plunging flows in bedrock rivers.\\

\noindent Here, we employ high-resolution numerical simulations to investigate the impact of constriction width and flow rate on the development of plunging flows.  Our central goal is to develop a numerical model that can predict the plunging flows observed in a laboratory flume experiment by \cite{hunt_experiments_2018} and then exploring aspects of flow dynamics that were not possible using the original laboratory observations. We address three central research questions: (1) Can subcritical plunging flows be reproduced in a numerical model? (2) What are the spatial and temporal characteristics of plunging flow structures? (3) Do numerical simulations support or challenge existing hypotheses on plunging flow formation? We document for the first time the mean and turbulent flow structures of plunging flows at high resolution to characterize the processes driving plunging flow formation.\\

\section{Experimental Design and Methods}
\subsection{Geometry and Flow Properties of the Prototype}
Our LES simulations were informed by the laboratory experiments of \cite{hunt_experiments_2018} who examined the effect of channel constriction on plunging flows in bedrock canyons.  The experiments were motivated by field observations of plunging flows through CPWs in the Fraser Canyon, a 375 km long bedrock reach of the Fraser River, British Columbia, Canada \parencite{hurson_abundance_2025,venditti_struggles_2020, venditti_flow_2014}. Morphological analysis of CPWs in the Fraser Canyon revealed that natural lateral constrictions typically ranged from 29\% to 51\% \parencite{venditti_struggles_2020,kusack_experiments_2024}.\\

\noindent The experimental flume was 4.8 m long, and 0.45 m deep, equipped with an adjustable sidewall to allow for variable channel widths (Figure \ref{Fig1}a). Two primary channel width configurations (0.1 m and 0.2 m) were tested in the \cite{hunt_experiments_2018} experiments, but all our CFD simulations were made with a 0.2 m wide channel. The flume setup included a 2.1 m entrance reach with a flat bed, followed by a 1.5 m long pool section with a constriction at the top of the pool, and a 1.2 m flat exit reach.   Lateral constrictions in the experiments were created ranging from 20\% to 50\% of the channel width (Figure \ref{Fig1}a).  Our numerical simulations aimed to replicate the geometry of the laboratory experiments, including the pool entrance slope and ratio of upstream flow depth to pool flow depth (i.e. 1.65).  The values for upstream flow depth, pool flow depth, and entrance velocity followed the experimental design of \cite{hunt_experiments_2018} (Table~\ref{Table1}).

\newpage
\begin{table}[htbp]
\centering
\caption{Geometric and Hydraulic Properties of a Prototype CPW Morphology Used in \cite{hunt_experiments_2018} Flume Experiments Informing the LES Model}
\label{Table1}

\begin{tabularx}{\textwidth}{
    >{\raggedright\arraybackslash}X
    >{\centering\arraybackslash}p{0.18\textwidth}
    >{\centering\arraybackslash}p{0.30\textwidth}
}
\toprule
\textbf{Property} &
\textbf{Prototype} &
\textbf{Laboratory Setup/LES Model} \\
\midrule

Constriction (\%) &
51 &
20--50 \\

Pool entrance slope ($^\circ$) &
12 &
5--15 \\

Upstream flow depth, $d_1$ (m) &
10 &
0.08 \\

Pool flow depth, $d_2$ (m) &
16.5 &
0.13 \\

$d_1/d_2$ &
1.65 &
1.65 \\

Entrance flow velocity, $V_1$ (m\,s$^{-1}$) &
2.3 &
0.25--0.45 \\

Channel width (m) &
40--80 &
0.2 \\

\bottomrule
\end{tabularx}
\end{table}

\subsection{CFD Model Structure and Conceptualization}
Our numerical modelling used OpenFOAM, open-source CFD software written in C++ \parencite{weller_tensorial_1998}. CFD models are solved by applying continuum equations, formulated within a tensorial framework, to cells of arbitrary shape using object-oriented programming (OOP) techniques and the finite volume method (FVM; \cite{bayon_performance_2016, mcdonald_computation_1971}). The OpenFOAM environment offers several significant benefits, including the flexibility to modify the code for further development, the availability of multiple solvers for a wide range of applications in continuum mechanics, and parallelization through the decomposition of the computational domain into subdomains, which reduces computational costs. The CFD model used in this study employs the FVM discretization \parencite{wang_model_2015} and a coupled Pressure Implicit Splitting of Operators (PISO) with Semi-Implicit Method for Pressure Linked Equations (SIMPLE) algorithm, called PimpleFOAM, to solve the transient incompressible Newtonian fluids. The momentum equation is solved using the LES technique. The LES model is three-dimensional, unsteady, and resolves anisotropic turbulence from large-scale motions (large eddies) to small-scale motions (subgrid scale), where the stresses are modeled in the sub-grid scales (SGS). To minimize the simulated time in this study, the computational domain is decomposed and parallelized across 10 processors of a supercomputer with 20 cores, each operating at 2.2 GHz and equipped with 32 GB of RAM. The Open Message Passing Interface (MPI) is used to communicate between processors \parencite{alvarez_detached_2017,alvarez_eddy-resolving_2021}. With the above specifications, a single simulation took approximately five hours to complete its calculations.

\subsubsection{Governing Equations}
The continuity and Navier–Stokes equations are solved using LES, which relies on a filtered form of the Navier–Stokes equations to resolve large-scale turbulent structures. The filtered governing continuity and momentum equations are:

\begin{equation}
\frac{\partial \overline{u}_i}{\partial x_i}=0
\label{eq:1}
\end{equation}

\begin{equation}
\frac{\partial \overline{u}_i}{\partial t}
+
\frac{\partial \left(\overline{u}_i\overline{u}_j\right)}
{\partial x_j}
=
-\frac{1}{\rho}
\frac{\partial \overline{p}}{\partial x_i}
+
\frac{\partial}{\partial x_j}
\left[
\left(\nu+\nu_t\right)
\left(
\frac{\partial \overline{u}_i}{\partial x_j}
+
\frac{\partial \overline{u}_j}{\partial x_i}
\right)
\right]
\label{eq:2}
\end{equation}

\noindent Here, $\overline{u}_i$ denotes the filtered velocity in the $i$-th direction 
($i =1,2,3$, corresponding to the Cartesian coordinates $x,y,z$), 
$\overline{p}$ is the filtered pressure, $\nu$ the molecular viscosity, and 
$\nu_t$ the eddy viscosity introduced by the sub grid-scale (SGS) model. 
Equation \ref{eq:1} enforces incompressibility by requiring that the divergence of the 
filtered velocity (i.e., $\partial \overline{u}_i/\partial x_i$) vanishes. 
In Equation \ref{eq:2}, each term of the filtered momentum equation has a distinct 
physical interpretation. The first term, 
$\partial \overline{u}_i/\partial t$, represents the local acceleration of 
the velocity component in the $i$-th direction. The nonlinear convective 
transport term, 
$\partial(\overline{u}_i\overline{u}_j)/\partial x_j$, describes momentum 
exchange due to interactions among velocity components. Here, 
$\overline{u}_j$ is the filtered velocity in the $j$-th direction 
($j=1,2,3$), so this term accounts for how momentum in direction $i$ is 
advected by flow components in all three spatial directions. On the 
right-hand side, the pressure gradient term, 
$-\partial \overline{p}/\rho\partial x_i$, represents the driving force due 
to pressure differences. The final term represents momentum diffusion due 
to both molecular viscosity $\nu$ and SGS contributions through the eddy 
viscosity $\nu_t$, which transfers energy from resolved eddies to unresolved 
scales.\\

\noindent The SGS stress tensor is modeled using the one-equation eddy-viscosity SGS 
model, denoted here by the model name \texttt{kEqn}, which solves a transport 
equation for SGS kinetic energy ($k_{\mathrm{sgs}}$) 
\parencite{greenshields_openfoam_2018-3,yoshizawa_statistical_1986}. This formulation avoids the need for 
ad-hoc damping functions because the eddy viscosity coefficient itself 
provides a natural dissipative mechanism, making the approach suitable for 
laboratory-scale flows \parencite{horiuti_large_1985,taghinia_one-equation_2015}. In contrast, 
damping functions in other SGS models are typically used to dissipate 
unresolved turbulent energy, prevent artificial amplification of resolved 
scales, and enhance numerical stability \parencite{taghinia_one-equation_2015}. In this 
study, the \texttt{PimpleFoam} solver is used, which enforces mass conservation 
(Equation \ref{eq:1}) through a pressure--velocity coupling loop and updates the 
velocity field to satisfy momentum conservation (Equation \ref{eq:2}) at each time 
step \parencite{greenshields_openfoam_2018}.\\

\subsubsection{Initial and Boundary Conditions}

\begin{figure}[htbp]
    \centering
    \includegraphics[width=\textwidth]{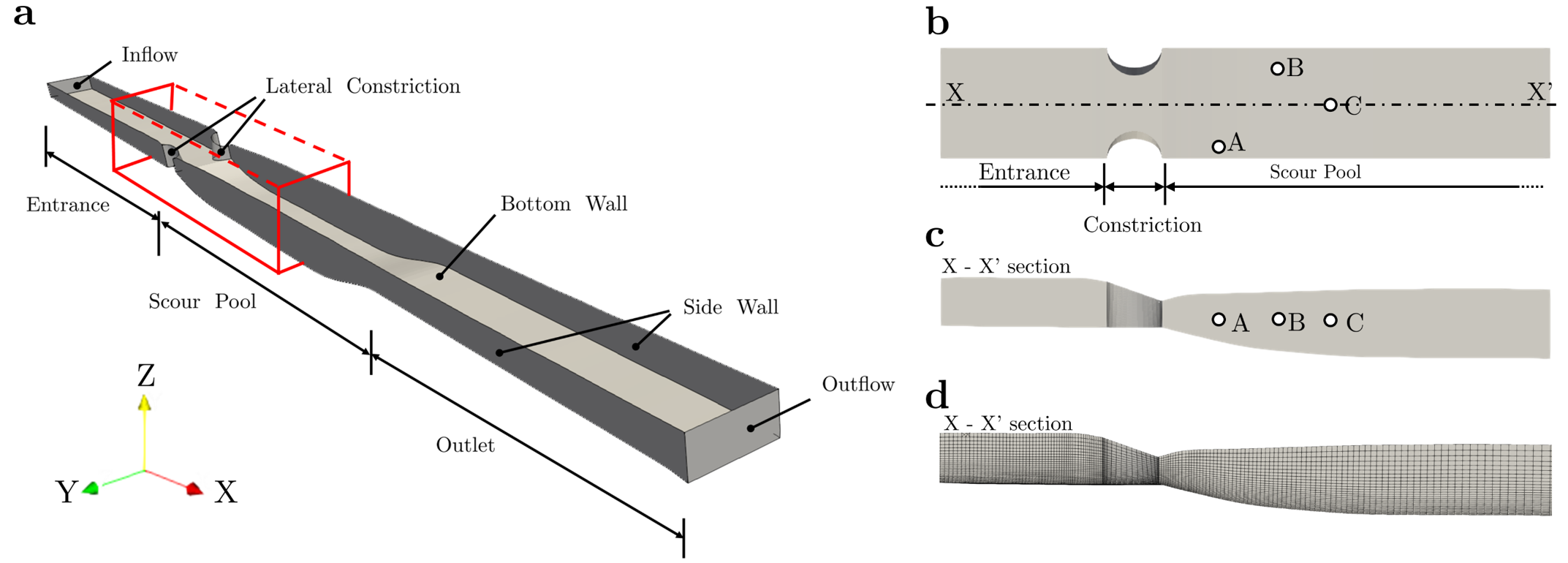}
    \vspace{-25pt}
\caption{LES computational domain and boundary conditions. (a) Three-dimensional (3D) view of the full channel geometry, illustrating key features including the entrance, lateral constriction, scour pool, and outlet. (b) Plan view of the highlighted region in panel a, showing the entrance, constriction, and scour pool. (c) Front view along the centerline X–X' (panel a). (d) Computational mesh from front view, showing mesh resolution from the entrance through the constriction and scour region. A, B, and C mark locations where we examined the oscillation patterns in turbulent flow field.}
    \label{Fig1}
\end{figure}

\noindent The laboratory setup featured a channel bed roughened with sand to simulate natural rock boundaries (Figure \ref{Fig1}). Accordingly, the bottom and side walls of the computational domain were assigned no-slip boundary conditions with an integrated rough wall function. Wall roughness was modeled as a constraint on turbulent viscosity ($\nu_t$), using a roughness height defined by $z_0 = 0.033k_s$ \parencite{garcia_sedimentation_2008}. The water surface was set as a free-slip boundary. At the inlet, initial and boundary conditions were specified according to a range of flow discharge, lateral constriction, and inflow velocity values (Table \ref{Table 2}). Our inlet velocities differ somewhat from those reported in \cite{hunt_experiments_2018}, because the values presented in their manuscript correspond to measurements taken at a specified downstream location within the channel rather than at the actual inlet. For the present simulations, we used inlet velocity values extracted from Hunt’s original experimental dataset at the actual channel inlet, which corresponds to our model’s upstream boundary.\\

\begin{table}[htbp]
\centering
\caption{Initial and boundary conditions used for the LES numerical model across the simulation scenarios.}
\label{Table 2}

\begin{tabularx}{\textwidth}{
    >{\centering\arraybackslash}p{0.12\textwidth}
    >{\centering\arraybackslash}p{0.16\textwidth}
    >{\centering\arraybackslash}p{0.17\textwidth}
    >{\centering\arraybackslash}p{0.22\textwidth}
    >{\centering\arraybackslash}X
}
\toprule
\textbf{Simulation ID} &
\textbf{Channel Width (m)} &
\textbf{Discharge (m$^3$\,s$^{-1}$)} &
\textbf{Lateral Constriction (\%)} &
\textbf{Velocity at Inflow, $U_{\mathrm{in}}$ (m\,s$^{-1}$)} \\
\midrule
LF0 & 0.2032 & 0.003 & No Constriction & 0.591 \\
C16 & 0.2032 & 0.003 & 20 & 0.470 \\
C17 & 0.2032 & 0.003 & 30 & 0.428 \\
C18 & 0.2032 & 0.003 & 35 & 0.385 \\
C19 & 0.2032 & 0.003 & 40 & 0.374 \\
C20 & 0.2032 & 0.003 & 50 & 0.319 \\
HF0 & 0.2032 & 0.006 & No Constriction & 0.680 \\
C21 & 0.2032 & 0.006 & 20 & 0.598 \\
C22 & 0.2032 & 0.006 & 30 & 0.421 \\
C23 & 0.2032 & 0.006 & 35 & 0.402 \\
C24 & 0.2032 & 0.006 & 40 & 0.382 \\
C25 & 0.2032 & 0.006 & 50 & 0.344 \\
\bottomrule
\end{tabularx}
\end{table}

\subsubsection{Computational Domain}

\begin{figure}[htbp]
    \centering
    \includegraphics[width=\textwidth]{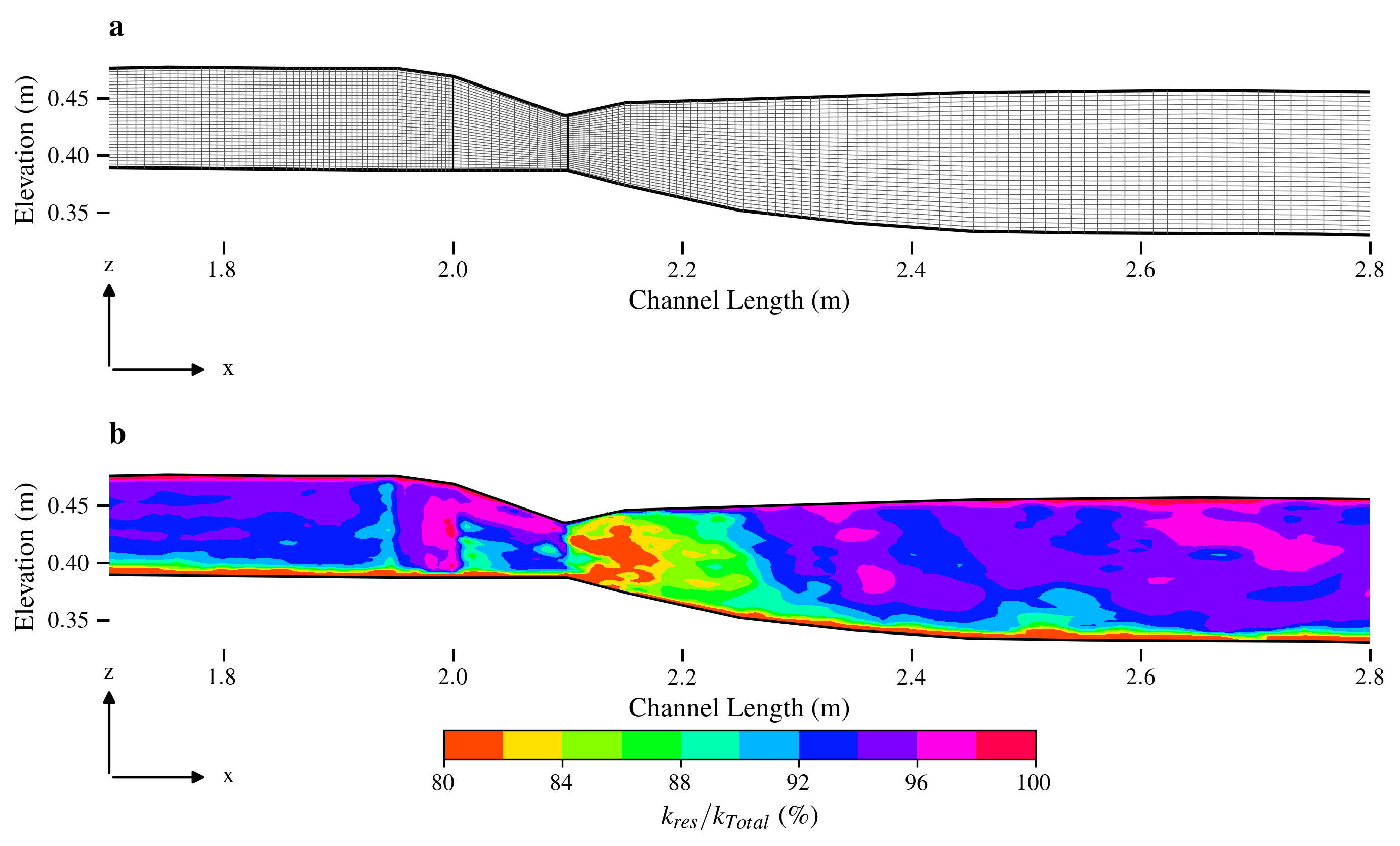}
    \vspace{-25pt}
\caption{(a) Selected adaptive resolution of the computational domain and (b) fraction of resolved Turbulent Kinetic Energy $k_{\mathrm{res}}/k_{\mathrm{total}}$ with the selected adaptive mesh resolution at 30s of simulation time for the C25 case.}
    \label{Fig2}
\end{figure}

The computational domain was designed using a hexagonal grid mesh for spatial discretization. The domain includes 472,500 cells arranged into three blocks: the entrance, lateral constriction, and scour pool/outlet. The mesh is structured with 315 cells in the streamwise ($x$) direction, 50 in the cross-stream ($y$), and 30 in the vertical ($z$). Selective mesh refinement was employed to resolve key flow features (Figure \ref{Fig2}a). Finer cells were made near the channel bed and gradually transitioned to coarser cells toward the free surface, allowing accurate resolution of near-bed shear and velocity gradients. Additional refinement was implemented at the lateral constriction and the upstream channel section. The entrance region, characterized by lower velocities and turbulence levels, and hence smaller integral length scales, required higher resolution \parencite{seifollahi_moghadam_evaluation_2021}. As flow entered the constriction, it accelerated due to a reduced cross-sectional area \parencite{holzmann_mathematics_2016}, forming a high-velocity jet in the scour pool \parencite{aider2007large, hunt_experiments_2018}. This generated intense turbulence and large-scale eddies \parencite{aider2007large,bayon_performance_2016}, allowing for a coarser mesh downstream where integral length scales increased \parencite{seifollahi_moghadam_evaluation_2021}.\\

\noindent The grid cell resolution of the computational domain was evaluated using the ratio of resolved turbulent kinetic energy ($k_{\mathrm{res}}$) to the total turbulent kinetic energy ($k_{\mathrm{total}}$) in the LES model. This ratio serves as a quantitative metric for assessing whether the grid adequately resolves turbulent structures. The total turbulent kinetic energy can be expressed as:

\begin{equation}
k_{\mathrm{total}} = k_{\mathrm{res}} + k_{\mathrm{sgs}}
\label{eq:total_tke}
\end{equation}

\noindent where $k_{\mathrm{sgs}}$ is the sub grid scale (SGS) turbulent kinetic energy. The resolved kinetic energy, $k_{\mathrm{res}}$, is calculated as

\begin{equation}
k_{\mathrm{res}} =
\frac{1}{2}
\left(
{u'}^2 + {v'}^2 + {w'}^2
\right)
\label{eq:resolved_tke}
\end{equation}

\noindent Here, $u'$, $v'$, and $w'$ are the fluctuating components of the velocity in the streamwise ($x$), lateral ($y$), and vertical ($z$) directions, respectively. These are defined as deviations from the mean velocities $u' = u - \overline{u}$, $v' = v - \overline{v}$, and $w' = w - \overline{w}$, with $u$, $v$, and $w$ denoting the instantaneous velocities. To ensure adequate spatial resolution, it is generally recommended that at least 80\% of the total turbulent kinetic energy be resolved (i.e., $k_{\mathrm{res}}/k_{\mathrm{total}} \geq 80\%$; \cite{grenouilloux_use_2022} \cite{pope_turbulent_2000}). The ratio $k_{\mathrm{res}}/k_{\mathrm{total}}$ was computed at each location using the one-equation SGS model for $k_{\mathrm{sgs}}$ \parencite{feng_comparisons_2021,yoshizawa_statistical_1986}. We found that $k_{\mathrm{res}}/k_{\mathrm{total}}$ ranged from 92--98\% upstream and at the lateral constriction, and was 80--99\% in the pool. Slightly lower values near the bed ($\sim$80\%) can be attributed to boundary layer effects. Overall, $k_{\mathrm{res}}/k_{\mathrm{total}} \geq 80\%$ across all regions (Figure \ref{Fig2}).\\

\noindent The selection of a suitable time step for the LES model was guided by the Courant--Friedrichs--Lewy (CFL) condition \parencite{courant_uber_1928}, which indicates numerical stability and convergence. The CFL condition quantifies how far a fluid particle, moving at velocity $U$, travels during a time interval $\Delta t$ relative to the local grid spacing $\Delta x$ as

\begin{equation}
\mathrm{CFL}=\frac{U\Delta t}{\Delta x}.
\label{eq:cfl}
\end{equation}

\noindent For computational stability and to improve the quality of the CFD simulation, the CFL number is generally limited to $<1.0$. In the case of arbitrarily shaped 3D cells, CFL is computed based on the smallest characteristic cell length and the maximum local velocity to ensure that the fluid does not traverse more than one cell per time step calculated as

\begin{equation}
\mathrm{CFL}=
\frac{1}{2}\Delta t
\left(
\frac{\sum_f \left| \mathbf{U}_f \cdot \hat{\mathbf{n}}_f \right| A_f}{V_p}
\right)
\label{eq:cfl_3d}
\end{equation}

\noindent where $\mathbf{U}_f$ is the velocity magnitude at the cell face, $\hat{\mathbf{n}}_f$ is unit normal vector to the face, $A_f$ is the surface area of the cell face, and $V_p$ is the volume of the computational cell. In fixed-time-step applications, OpenFOAM performs an internal calculation and tracking of the CFL number during the simulation. We selected the time step separately for each simulation case. This was necessary because the maximum velocity varied between cases. For all simulations, the time step was chosen so that $\mathrm{CFL}<1.0$ throughout the flow field.

\subsubsection{Resolved Stress Computation Using the LES Model}
The filtered momentum equation (Equation \ref{eq:2}) can be conveniently rewritten in consideration of the resolved and modeled SGS stress contributions as

\begin{equation}
\frac{\partial \overline{u}_i}{\partial t}
+
\frac{\partial \left(\overline{u}_i\overline{u}_j\right)}
{\partial x_j}
=
-\frac{1}{\rho}
\frac{\partial \overline{p}}{\partial x_i}
+
\frac{\partial}{\partial x_j}
\left(
\nu
\frac{\partial \overline{u}_i}{\partial x_j}
\right)
-
\frac{\partial \tau_{ij}}{\partial x_j}
\label{eq:filtered_momentum_sgs}
\end{equation}

\noindent where $\tau_{ij}=\overline{u_i u_j}-\overline{u}_i\overline{u}_j$ represents the SGS stress tensor, responsible for the momentum transfer by the unresolved turbulent fluctuations. This is derived from Equation (\ref{eq:2}) using the decomposition of the overall diffusive term

\begin{equation}
\frac{\partial}{\partial x_j}
\left[
\left(\nu+\nu_t\right)
\left(
\frac{\partial \overline{u}_i}{\partial x_j}
+
\frac{\partial \overline{u}_j}{\partial x_i}
\right)
\right]
\label{eq:overall_diffusive_term}
\end{equation}

\noindent into a molecular viscosity term and a modeled SGS stress divergence using the product rule. The explicit separation permits turbulence effects to be directly examined. The deviatoric part of the SGS stress tensor is formulated via an eddy-viscosity assumption. For the one-equation SGS model, $\tau_{ij}$ is expressed as

\begin{equation}
\tau_{ij}=2\nu_t S_{ij},
\qquad
S_{ij}
=
\frac{1}{2}
\left(
\frac{\partial \overline{u}_i}{\partial x_j}
+
\frac{\partial \overline{u}_j}{\partial x_i}
\right)
\label{eq:sgs_stress}
\end{equation}

\noindent where $\nu_t=C_k\sqrt{k_{\mathrm{sgs}}\Delta}$ is the eddy viscosity, $k_{\mathrm{sgs}}$ is the SGS turbulent kinetic energy, $\Delta$ is the filter width, and $C_k$ is a model constant. This formulation links the SGS stresses to the local strain rate and energy content of unresolved turbulence. Therefore, the total of the turbulent stress tensor can be expressed as $\tau_{ij}=\overline{u_i u_j}-\overline{u}_i\overline{u}_j$ where the first term is the resolved component and the second is the modeled SGS stress. This LES configuration resolves over 80\% of the turbulent kinetic energy, ensuring that the grid resolution is sufficient to capture the dominant momentum transport processes and anisotropy of plunging flows.

\subsection{Methodology for Model Validation}

The validation of the numerical simulations was performed using data from the laboratory experiments that employed dye tracing to visualize flow structure under varying constriction conditions. \cite{hunt_experiments_2018} developed a method to calculate velocity form dye traces but applied the methods to only three discrete locations for each constriction case: upstream of the constriction, at the constriction, and downstream of the constriction. These limited data points were insufficient to validate the numerical model or to fully characterize the spatial variability of the flow field. To address this limitation, we digitized dye traces using the WebPlotDigitizer tool \parencite{rohatgi_webplotdigitizer_2023} and generated velocity profiles along the channel centerline following the method in \cite{hunt_experiments_2018}. This produced velocity profiles that could be directly compared to the numerical simulation.  We assumed that the dye traces represented the centerline velocities, corresponding to the maximum velocity along the flow.

\subsection{Evaluating Mean Flow Characteristics}

To examine the mean flow field, we explored mean velocity distributions, velocity profile inversion metrics, near-bed shear stress distributions, and streamlines. The flow regime was classified according to the Froude number calculated as

\begin{equation}
Fr = U\sqrt{gh}
\label{eq:froude_number}
\end{equation}

\noindent where $U$ represents the local flow velocity, $g$ is the gravitational acceleration, and $h$ the flow depth. Observations in the Fraser Canyon indicate that plunging flows are a subcritical flow phenomenon in terms of the Froude number \parencite{venditti_flow_2014,wright_covariation_2022}, but laboratory experiments demonstrate that flow transitions from subcritical to transcritical conditions ($0.9 < Fr < 1.1$) through constrictions \parencite{hunt_experiments_2018}. We calculated $Fr$ at three key positions: upstream of the constriction, at the constriction, and immediately downstream. Both low- and high-flow scenarios were evaluated to assess how constriction width influences subcritical and transcritical transitions.

\noindent The strength of velocity inversion was quantified by the relative position of the maximum velocity below the water surface, calculated as:

\begin{equation}
\delta_{\max} = \left(
1-\frac{h_{\mathrm{maxVel}}}{h}\times100%
\right)
\label{eq:velocity_inversion}
\end{equation}

\noindent where $h_{\mathrm{maxVel}}$ is the vertical distance from the bed to the point of maximum velocity. and $h$ is the local flow depth. In the experiments, \cite{hunt_experiments_2018} used maximum dye concentration to give the highest velocity and in the simulations, we used maximum velocity. This metric indicates how far the maximum velocity is below the water surface. We classified velocity inversions following \cite{hunt_experiments_2018} as: no inversion ($\delta_{\max} < 30\%$), weak ($30\% < \delta_{\max} < 45\%$), moderate ($45\% < \delta_{\max} < 60\%$), and strong ($\delta_{\max} > 60\%$).\\

\noindent Given the influence of velocity structure on sediment transport, particular attention was paid to the distribution of bed shear stress, which governs erosional processes and the formation of scour features. Velocity inversions enhance streamwise velocities near the bed, thereby reinforcing shear stress and the potential for erosion and vertical incision. Instantaneous shear stress was computed as

\begin{equation}
\tau_{\mathrm{bed}} =\rho\nu_t\frac{\partial u}{\partial z}
\label{eq12}
\end{equation}

\noindent where $\nu_t$ is the eddy viscosity, and the velocity gradient $\partial u/\partial z$ was derived from LES output in order to represent local shear effects at high spatial fidelity. OpenFOAM reports wall shear stress in kinematic units, but we multiplied the computed values by fluid density, $\rho$, to express wall shear stress in Pascals.\\

\noindent To facilitate better visualization of the spatial structure of the plunging flows and associated recirculation zones, post-processing streamlines were generated using the Stream Tracer module of ParaView (OpenFOAM). Streamlines were calculated by integrating the steady-state velocity field using a collection of velocity points with numerical solvers (i.e. fourth-order Runge--Kutta). These visualizations give qualitative data on curvature of flow, upwelling-downwelling, and convergence patterns that agree with plunging flow, upwelling-downwelling, and convergence patterns that agree with plunging flow mechanisms and secondary circulation phenomena.

\subsection{Assessment of Turbulent Flow Fields}
The turbulent flow field is examined by exploring the distributions of turbulence intensities (root-mean-square velocities), Reynolds stress, and turbulent kinetic energy. Root mean square streamwise and vertical velocities were calculated as

\begin{equation}
u_{\mathrm{rms}}
=
\left[
\frac{1}{n}
\sum
\left(
u-\overline{u}
\right)^2
\right]^{0.5}
\label{eq:u_rms}
\end{equation}

\begin{equation}
w_{\mathrm{rms}}
=
\left[
\frac{1}{n}
\sum
\left(
w-\overline{w}
\right)^2
\right]^{0.5}
\label{eq:w_rms}
\end{equation}

\noindent The resolved Reynolds stress tensor was examined to assess the directionality of the turbulent energy distribution and to validate the LES model. OpenFOAM’s \texttt{FieldAverage} function was utilized to approximate the time averaged product of the velocity components as $\tau_{ij}=\overline{u_i u_j}-\overline{u}_i\overline{u}_j$. The resulting symmetric tensor includes three normal stresses ($u'^2$, $v'^2$, $w'^2$), and three shear stresses ($u'v'$, $u'w'$, $v'w'$). We focused on the downstream and vertical component of the Reynolds shear stress tensor which dominates fluid stress and is calculated as

\begin{equation}
\tau_{uw}=\rho\overline{u'w'}
\label{eq:reynolds_shear_stress}
\end{equation}

\noindent The resolved turbulent kinetic energy was calculated from the normal components using Equation 4. The shear components were further analyzed to examine directional momentum exchange and turbulent mixing in the flow.\\

\noindent We examined the magnitude of turbulence production associated with vertical shear to identify regions where mean-flow kinetic energy is converted into turbulent kinetic energy. For the present two-dimensional analysis, the dominant shear-production term was evaluated following as \cite{venditti2000spectral}, as

\begin{equation}
P=-\overline{u'w'}\frac{\partial U}{\partial z}
\label{eq:turbulence_production}
\end{equation}

\noindent where $P$ is the turbulence production due to vertical shear, $U$ is the time-averaged streamwise velocity, $z$ is the vertical coordinate, and $\overline{u'w'}$ is the Reynolds shear stress component. Positive values of $P$ indicate that kinetic energy is transferred from the mean flow to turbulent fluctuations, representing active turbulence generation. Negative values indicate localized reverse energy exchange, where turbulent motions transfer energy back toward the mean flow. Distributions of $P$ indicate how channel constriction and plunging-flow modify the vertical shear structure and the spatial distribution of turbulence generation within the pool.\\

\noindent To further evaluate turbulent momentum exchange, the eddy viscosity associated with vertical momentum transport ($\epsilon$) was computed as\\

\begin{equation}
\epsilon=
\frac{-\overline{u'w'}}
{\partial U/\partial z}
\label{eq:eddy_viscosity}
\end{equation}

\noindent following \cite{venditti2000spectral}. $\epsilon$ is a flow-dependent measure of how efficiently turbulent eddies mix momentum across vertical layers of the flow. Positive values of $\epsilon$ indicate stronger turbulent momentum exchange, whereas negative values indicate regions where the turbulent momentum flux locally opposes the mean velocity gradient. Turbulence production identifies where mean-flow energy is converted into turbulence. Eddy viscosity describes how strongly turbulent motions redistribute momentum.\\

\noindent Oscillatory flow behavior was examined using the continuous wavelet transform (CWT) of velocity magnitude time series. The CWT is a time-frequency analysis method that decomposes a signal into localized oscillatory components, allowing periodic features and coherent flow structures to be identified as they evolve in time. We examined three locations in the flow for the 50\% constriction case (C25): Location A at the center of an eddy formed downstream of the constriction on channel (Eddy 1; $x = 2.2$ m, $y = 0.18$ m, $z = 0.4$ m), Location B in an eddy formed downstream of the constriction on channel (Eddy 2; $x = 2.3$ m, $y = 0.036$ m, $z = 0.4$ m), and Location C at the pool center ($x = 2.4$ m, $y = 0.1$ m, $z = 0.4$ m) (Figure \ref{Fig1}). Velocity magnitude was extracted at these points and analyzed to determine dominant oscillation periods and their temporal variability.\\

\noindent We also employed spectral analysis to further examine the dominant oscillation frequencies at Locations A, B, and C. Velocity signals were detrended by subtracting the mean. A band-pass frequency range of 0.05--2.0\,Hz was then applied to retain the dominant oscillatory motions while reducing the influence of low-frequency drift and high-frequency noise. For each location, the power spectral density, $P(f)$, was computed from the detrended velocity time series. The spectrum was then expressed in variance-preserving form as $\left(fP(f)\right)/\sigma^2$ where $f$ is frequency, $P(f)$ is the power spectral density, and $\sigma^2$ is the variance of the detrended velocity signal \parencite{venditti2000spectral,venditti2005turbulent}. This normalization allows the spectra from different locations to be compared based on the relative contribution of each frequency band to the total velocity variance \parencite{kaimal_atmospheric_1994,panofsky_atmospheric_1984,soulsby_similarity_1977}. Peaks in the variance-preserving spectra were used to identify the dominant oscillation frequencies and the associated energetic eddy time scales.

\section{Results}
\subsection{Model Validation}

Under low-flow conditions, both observed and simulated velocities remain relatively low with limited variation across the constriction cases (Figure \ref{Fig3}a). Median velocity derived from dye tracing ranges from 0.23–0.26 m s\textsuperscript{-1}, while simulated median velocity falls between 0.19–0.26 m s\textsuperscript{-1}, indicating a modest bias in the CFD results. The simulations also show narrower interquartile ranges than the observations. This reduced spread reflects the smoothing effect of the LES model and the absence of measurement-related noise, which likely contributes to the stronger positive skewness in the observed data.\\ 

\noindent Under high-flow conditions, both observed and simulated velocity distributions show a general increase in velocity with increasing constriction, consistent with the expected acceleration (Figure \ref{Fig3}b). Observed median velocities increased gradually from 0.40 to 0.42 m s\textsuperscript{-1} between 20\% and 35\% constriction, followed by a sharp rise to 0.73 m s\textsuperscript{-1} at 50\% constriction, and the widening interquartile ranges and strong positive skewness suggest intensified acceleration and turbulent fluctuations within the constricted flow section. However, these experimentally derived velocities should be interpreted with caution because dye-tracking techniques are subject to dispersion, uneven mixing, and tracking uncertainty, which may inflate the apparent variability. The simulated velocities remained nearly constant at approximately 0.50 m s\textsuperscript{-1} from 20\% to 40\% channel constriction, then increased sharply to 0.67 m s\textsuperscript{-1} at 50\% constriction. Relative to the observations, the simulations overpredict the median velocity at lower constrictions, particularly at 20\%, 30\% and 35\%, while they closely match the median at 40\% and slightly under predicts at 50\%. The interquartile ranges of the simulated distributions are generally narrower than those of the observations at 30–40\% constriction, suggesting reduced variability in the modeled results, whereas at 50\% constriction both observed and simulated distributions become much wider.

\begin{figure}[htbp]
    \centering
    \includegraphics[width=0.92\textwidth]{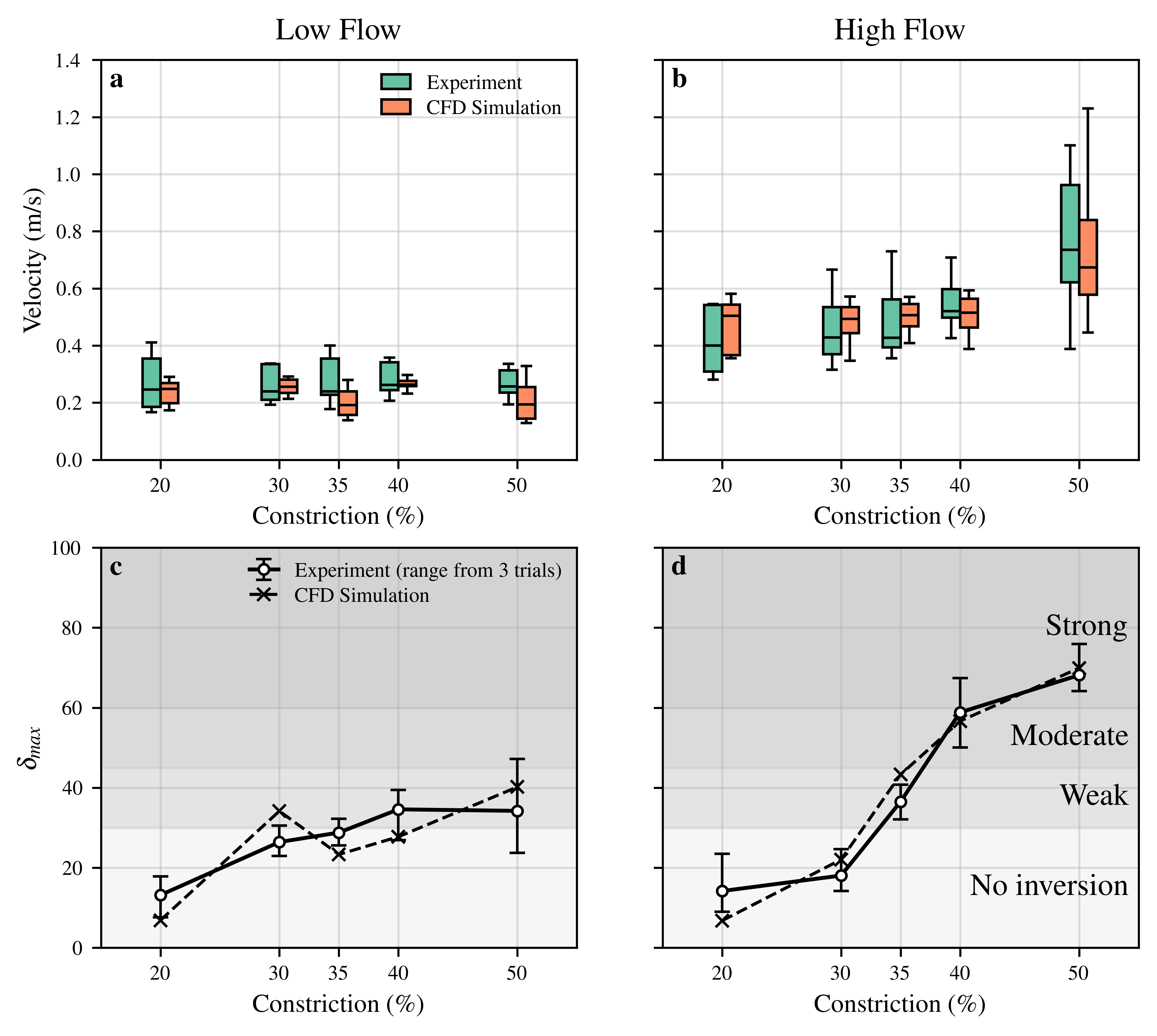}
\end{figure}

\clearpage

\noindent
\begin{minipage}{\textwidth}
    \captionof{figure}{Comparison of observed streamwise velocity calculated from dye tracing and CFD simulated streamwise velocity for a) low-flow and b) high-flow conditions. Inversion strength calculated as the relative depth of maximum velocity ($\delta_{max}$) for c) low-flow and d) high-flow.}
    \label{Fig3}
\end{minipage}

\noindent Further comparison between laboratory experiments and CFD simulations was performed to assess the model ability to reproduce velocity inversion strength behavior across varying lateral constrictions and discharge conditions. Under low-flow conditions, the CFD simulations reproduced the overall trend in inversion strength behavior but exhibited variable agreement across constrictions (Figure \ref{Fig3}c). At 40\% and 50\% constriction, the CFD predictions fell within the experimental uncertainty range. At 20\% and 35\% constriction, the simulated values are slightly lower than the experimental mean but remain broadly consistent with the observed trend. At 30\% constriction the simulations indicated weak inversion, although the experimental bounds suggest that no-inversion. \\

\noindent We calculated error metrics to assess model performance that included the root mean square error (RSME) and mean absolute error (MAE) (Table \ref{Table3}).  RSME and MAE are calculated from the residuals between the CFD-predicted inversion strength and the experimental mean inversion strength at each constriction level, where the experimental mean was obtained by averaging the three laboratory trials. Thus, these metrics quantify the difference between the model and the experiments rather than the variability within the experiments themselves. The experimental uncertainty range was defined by the lower and upper bounds of the three trials. Error metrics show RMSE and MAE values ranging from 5.46\% to 7.74\%, with the smallest error at 35\% (RMSE = 5.46\%) and the largest at 30\% (RMSE = 7.74\%). Across the low-flow constriction range, approximately 60\% of CFD predictions fell within the experimental uncertainty envelope (Table \ref{Table3}).\\

\noindent Under high-flow conditions, the CFD model showed markedly improved agreement with the laboratory data (Figure \ref{Fig3}d). The simulations captured the plunging velocity cores observed at higher constrictions and followed the experimental trend in relative depth more closely. RMSE and MAE values ranged from 1.76\% to 7.41\%, with the smallest error at 50\% (RMSE = 1.76\%) and the largest at 20\% (RMSE = 7.41\%) (Table \ref{Table3}). CFD predictions for 30\%, 40\%, and 50\% constrictions fell within the corresponding experimental uncertainty bounds. Because only one paired experimental observation was available per constriction, RMSE and MAE are identical in magnitude. Overall, the high-flow results show that the CFD model effectively captures the general pattern of velocity inversion, suggesting the dominant physical processes governing velocity inversion are represented by the CFD model.\\

\vspace{-28pt}
\begin{table}[htbp]
\centering
\caption{Statistical error metrics between laboratory experiment and simulations}

\label{Table3}

\begin{tabularx}{\textwidth}{
    >{\raggedright\arraybackslash}p{0.20\textwidth}
    >{\centering\arraybackslash}p{0.15\textwidth}
    >{\centering\arraybackslash}p{0.15\textwidth}
    >{\centering\arraybackslash}p{0.15\textwidth}
    >{\centering\arraybackslash}X
}
\toprule
\textbf{Flow Condition} &
\textbf{Model} &
\textbf{RMSE} &
\textbf{MAE} &
\textbf{Within Error Bounds} \\
\midrule

Low Flow  & C16 & 6.29 & 6.29 & No  \\
          & C17 & 7.74 & 7.74 & No  \\
          & C18 & 5.46 & 5.46 & No  \\
          & C19 & 6.85 & 6.85 & Yes \\
          & C20 & 5.97 & 5.97 & Yes \\

\midrule

High Flow & C21 & 7.41 & 7.41 & No  \\
          & C22 & 3.99 & 3.99 & Yes \\
          & C23 & 6.79 & 6.79 & No  \\
          & C24 & 2.17 & 2.17 & Yes \\
          & C25 & 1.76 & 1.76 & Yes \\

\bottomrule
\end{tabularx}
\end{table}
\newpage
\subsection{Mean Flow Characteristics}
\subsubsection{Hydraulic flow regime}
\begin{figure}[h!]
    \centering
    \includegraphics[width=1\textwidth]{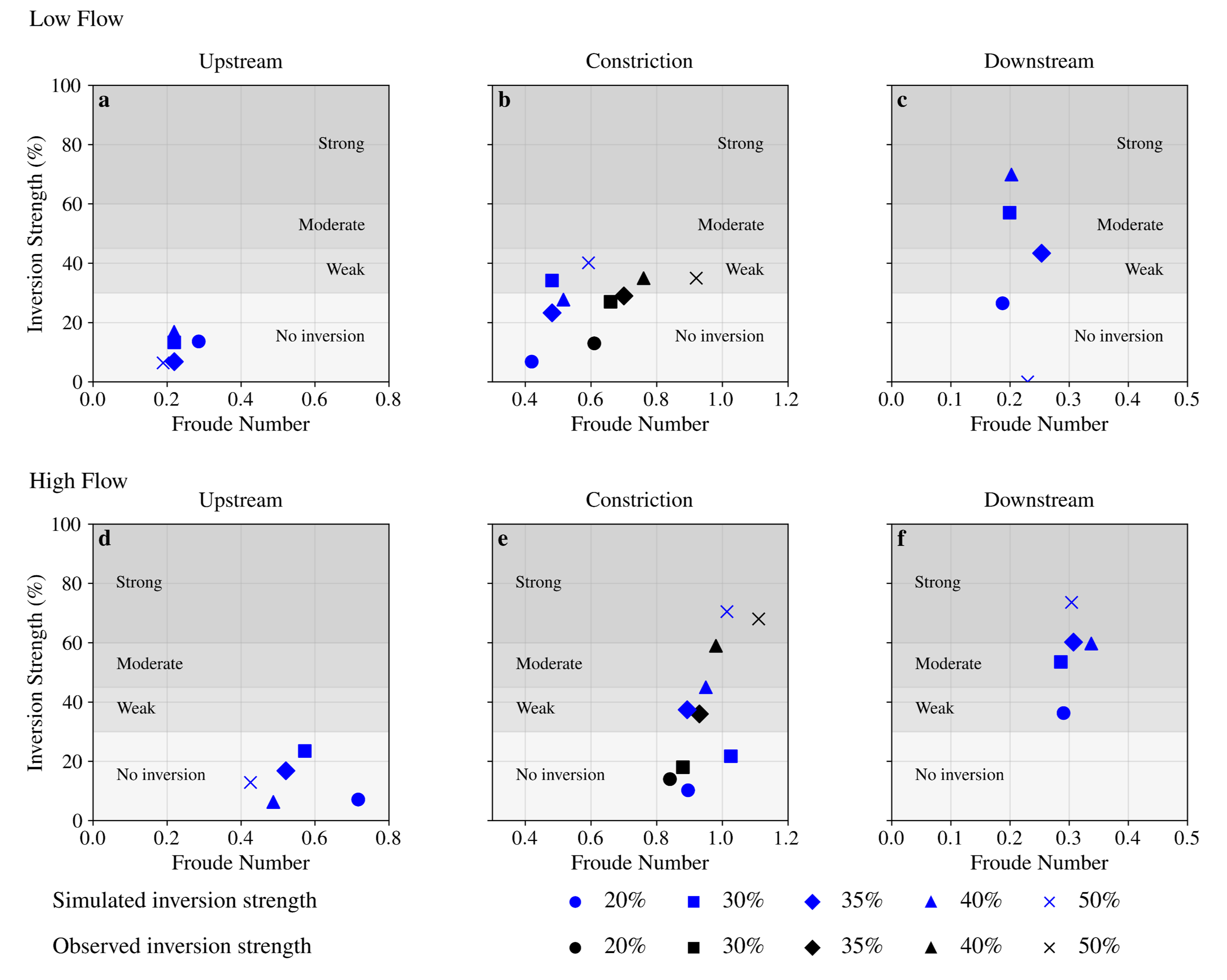}
    \caption{Comparison of Froude number ($Fr$), velocity, and inversion strength ($\delta_{max}$) under low-flow (a–c) and high-flow (d – f) conditions.}

    \label{Fig4}
\end{figure}

Under low-discharge conditions, simulated flow remained subcritical ($Fr < 1.0$) along the channel centerline, with systematic changes in velocity and inversion strength ($\delta_{\max}$) observed across the constriction. Upstream of the constriction, Froude numbers ranged from 0.18 to 0.28, confirming stable subcritical flow (Figure \ref{Fig4}a). Within the constriction, Froude number increased with narrowing channel width, reaching values of 0.41--0.59 and 0.62--0.92, respectively (Figure \ref{Fig4}b). Inversion strength also increased but remained within the weak category (30--45\%). Simulated inversion values (6.8--40\%) were generally lower than the observed values (13--35\%) for C16, C18, and C19, suggesting that the model underestimates inversion intensity for these cases. However, the model overestimates inversion intensity for C17 and C20. Downstream of the constriction, the flow decelerated, with Froude numbers falling below 0.3, reestablishing subcritical conditions (Figure \ref{Fig4}c). Inversion persisted for intermediate constrictions (30--40\%) but weakened at 50\%, indicating diminished plunging behavior. Overall, the model reproduced the qualitative trends in flow acceleration and inversion, though it underestimated their magnitudes.\\

\noindent At high discharge, both flow velocity and plunging behavior intensified. Upstream of the constriction, observed and simulated flows remained subcritical, with Froude numbers ranging from 0.38 to 0.56 and 0.43 to 0.72, respectively (Figure \ref{Fig4}d). These values exceeded those under low-flow conditions, reflecting increased momentum and reduced relative depth. Within the constriction, flow acceleration was evident, with observed Froude numbers increasing to 0.84 – 1.11 and simulated values to 0.89–1.03, approaching critical conditions (Figure \ref{Fig4}e). Inversion strength increased with constriction severity, ranging from 14 to 68\% in observations and 10 to 70\% in simulations. At 50\% constriction, inversion reached strong intensity, marking the onset of plunging-flow conditions. Downstream, flow decelerated and returned to a subcritical regime, with Froude numbers decreasing to 0.21 – 0.34 (Figure \ref{Fig4}f). Despite flow deceleration, inversion persisted for constrictions exceeding 35\%, with peak velocities occurring below 60\% of the flow depth. The numerical model effectively captured the observed patterns of acceleration, deceleration, and inversion persistence, though it slightly overestimated inversion magnitude at the highest constriction levels.\\

\vspace{-35pt}

\subsubsection{Mean flow field}
\begin{figure}[h!]
    \centering
    \includegraphics[width=0.99\textwidth]{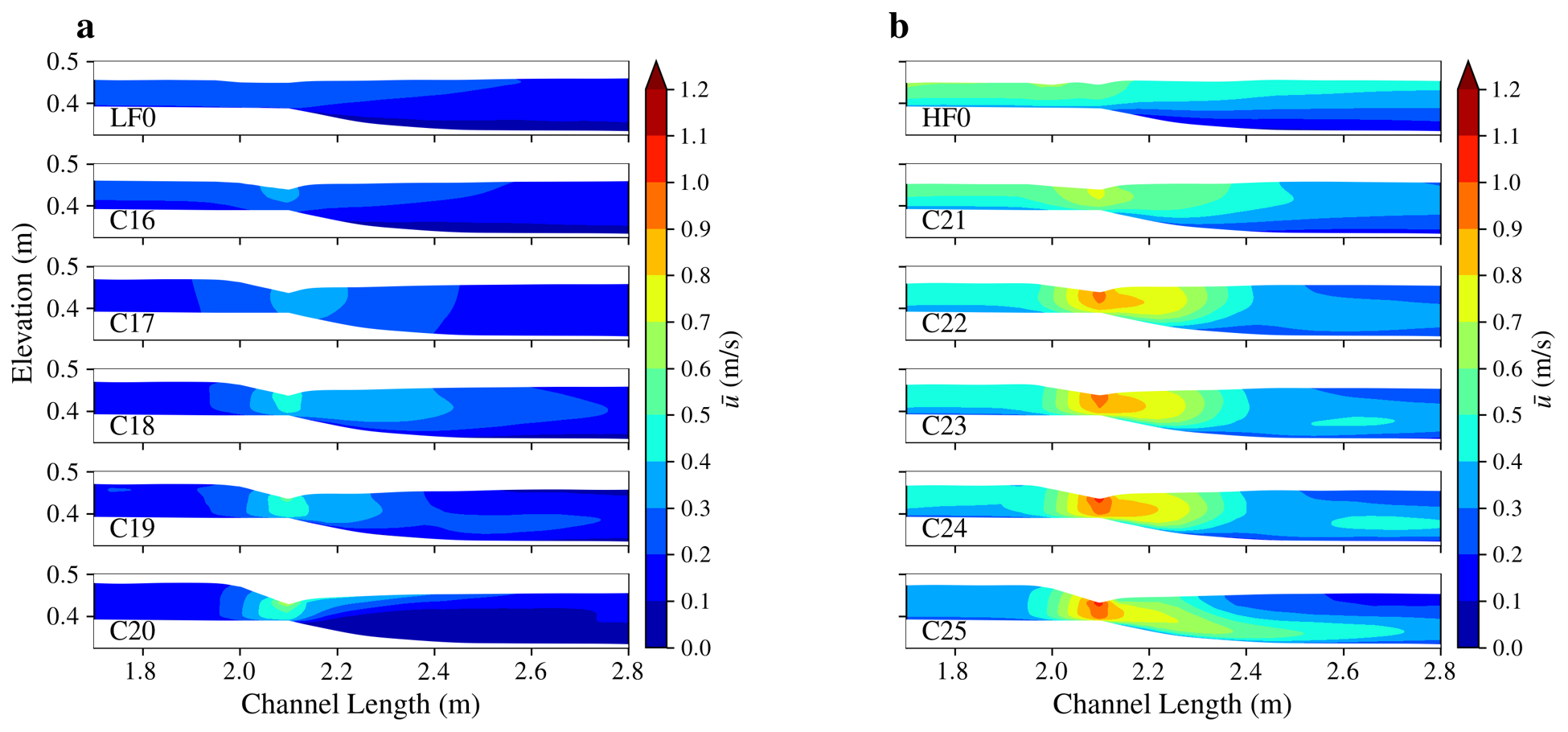}
    \caption{Mean velocity magnitude for (a) low flow and (b) high flow simulations.}
    \label{Fig5}
\end{figure}

Flow fields from low-flow simulations reveal distinct patterns in velocity as lateral constriction increases (Figure \ref{Fig5}a). For cases ranging from no lateral constriction (LF0) to 40\% lateral constriction (C19), maximum velocities shifted toward the channel bed as the degree of constriction increased. However, these conditions only produced weak velocity inversions, with  $\delta_{max}$ ranging from 30\% to 45\%. For the 50\% lateral constriction (C20), the inflow velocity was comparatively lower than in other low-flow scenarios, likely due to a backwater effect. This reduction occurred because higher downstream resistance in the more constricted cases caused an increase in upstream water level, which in turn lowered the inlet velocity to maintain the same discharge. As the system adjusts to increasing energy losses downstream, the inflow depth rises and velocity decreases according to fluid continuity. This reduction led to lateral velocity reversals immediately downstream of the constriction, attributed to the sudden narrowing of the channel and the associated geometric and flow rate changes. Low-flow conditions with increasing lateral constriction and decreasing inflow velocity appear insufficient to generate strong velocity inversions. Nevertheless, weak inversions remain possible under such conditions. \\

\noindent Different dynamics occurred in the high-flow scenario with increasing lateral constriction (Figure \ref{Fig5}b). Simulations indicate that velocities increased through the constriction, with higher velocities shifting toward the channel bed as the degree of constriction increased from 20\% (C21) to 50\% (C25). The velocity magnitude amplified with increasing constriction, accelerating the flow. Although 50\% constriction (C25) exhibited a slightly lower velocity compared to C21–C24.  Despite the higher flow rate and greater narrowing this condition produced water-level superelevation above the constriction. This phenomenon redirected flow downward as an impinging jet, subsequently creating a submerged hydraulic jump-like feature under subcritical conditions. Unlike a classical submerged hydraulic jumps formed from supercritical flow, this feature developed because the local depth at the constriction exceeded the optimal depth required for a free jump to be fully developed. These dynamics likely drive the high-velocity inversions observed in high-flow scenarios. 

\subsubsection{Elevation of maximum velocity and inversions}

\begin{figure}[h!]
    \centering
    \includegraphics[width=0.99\textwidth]{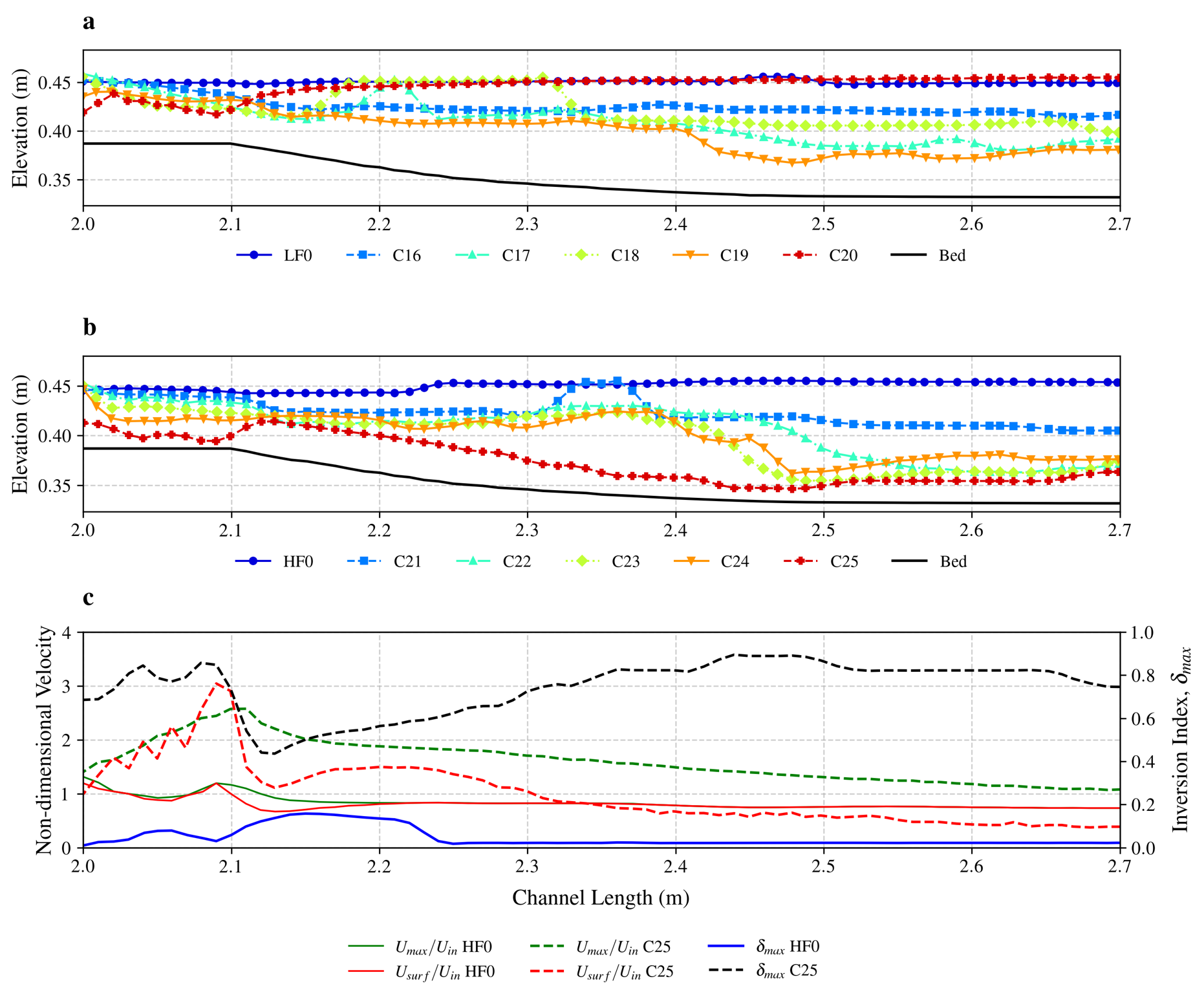}
    \vspace{-15pt}
    \caption{Maximum velocity pathways for a) low flow (LF0, C16-C20) and b) high flow (HF0, C21-C25). (c) Non-dimensional velocity and relative depth of maximum velocity ($\delta_{max}$) for 50\% constriction simulations at low flow (C20) and high flow (C25).}
    \label{Fig6}
\end{figure}

The maximum velocity pathways in the simulations were determined for both low and high flows. For the low flow cases, increasing constriction and decreasing inflow velocity pushed the maximum velocity pathways toward the bed for cases C16–C19 (Figure \ref{Fig6}a). However, in the 50\% constriction (C20), the thread of highest velocity stayed at the surface and the maximum velocity appeared similar to that in LF0 (no constriction). There is not a clear trend in the elevation of he maximum velocity pathway with constriction.  Among the constrictions, C19 (40\% constriction) exhibited the strongest velocity inversion. This contrasts with \cite{hunt_experiments_2018} who found the greatest inversion occurred for the 50\% constriction.\\

\noindent For the high-flow cases, the pathway of maximum velocity shows a clear downward shift as the degree of lateral constriction increases (Figure \ref{Fig6}b). In the unconstricted case, HF0, the maximum velocity remains close to the free surface throughout the pool, indicating no velocity inversion. With increasing constriction from C21 to C25, the maximum velocity pathway progressively moves closer to the bed, showing that the high-velocity core is displaced downward as the plunging-flow structure becomes stronger. The location where the maximum velocity begins to shift downward also moves upstream with increasing constriction. Farther downstream, beyond approximately x = 2.55m, the maximum velocity pathways become nearly constant for all high-flow cases. This suggests that after the initial adjustment zone downstream of the constriction, the flow reaches a more stable vertical structure within the pool. Overall, Figure \ref{Fig6}b shows that increasing constriction enhances the velocity inversion. We compared the ratios of the maximum velocity ($U_{max}$) and surface velocities ($U_{surf}$) to the $U_{in}$   (Figure \ref{Fig6}c). For the no constriction case (HF0), the $U_{max}$ /$U_{in}$  and $U_{surf}$ /$U_{in}$ are nearly constant along the channel, slightly >1 upstream of the constrictions and slightly < 1 downstream of the constriction. $U_{max}$ reveals no plunging flow.  For the high-flow, 50\% constriction (C25), $U_{max}$ /$U_{in}$  increased sharply  within the constriction (x = 2.1m), reaching values greater than 3, then remained > 1 over the pool, while $U_{surf}$ /$U_{in}$  increased in the constriction, then decreased downstream of the constriction (Figure \ref{Fig6}c). This indicates a strong downward redistribution of momentum and the formation of a pronounced plunging flow. For the high flow with 50\% constriction, the relative depth of maximum velocity $\delta_{max}$ also increased rapidly downstream of the constriction and stabilized around 0.8, confirming the downward shift of the velocity core and enhanced near-bed shear.

\subsubsection{Effect of velocity inversion on bed shear stress}

\begin{figure}[h!]
    \centering
    \includegraphics[width=\textwidth]{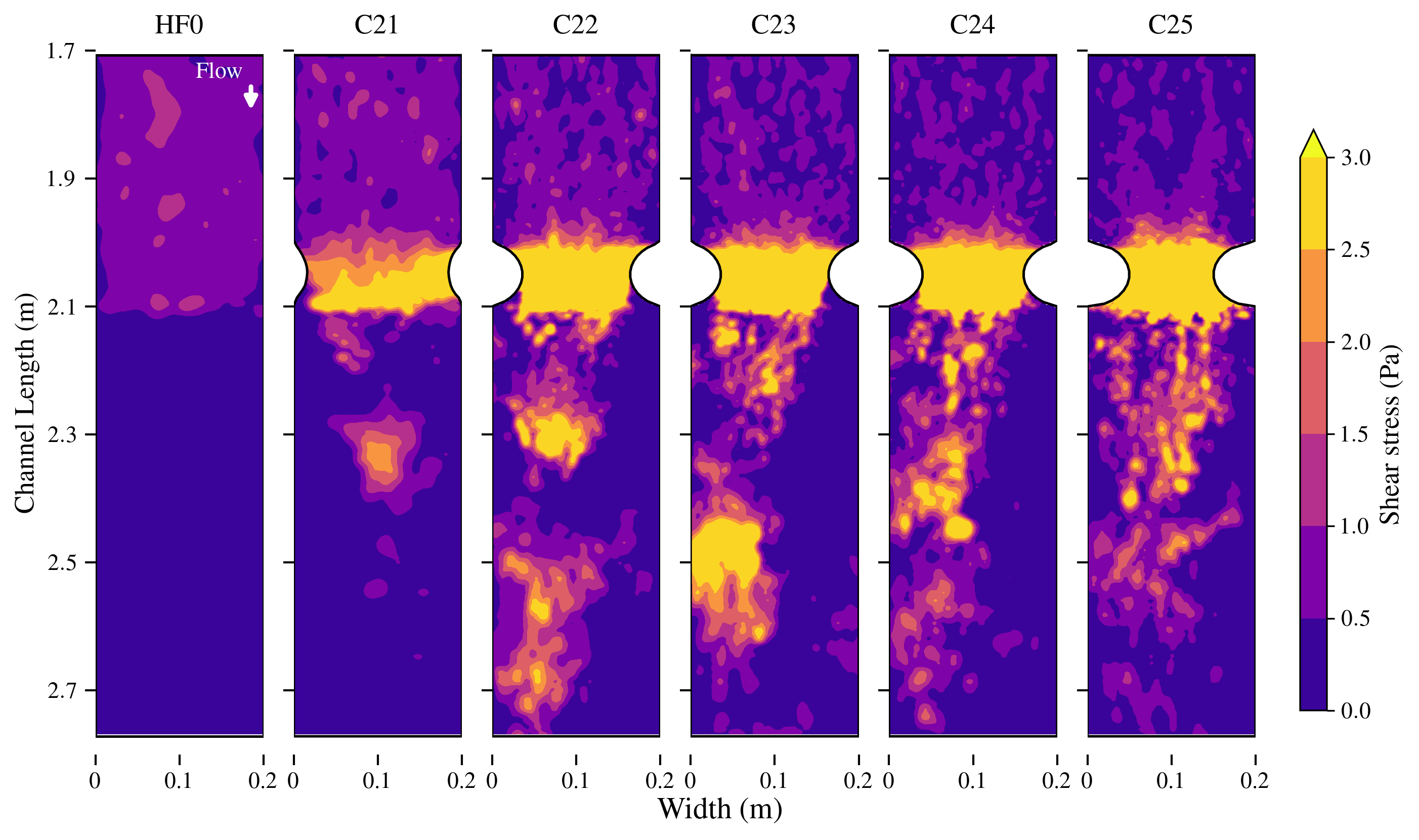}
    \caption{Instantaneous bed shear stress (Equation \ref{eq12}) magnitude for high flow scenario.}
    \label{Fig7}
\end{figure}

Vertical incision and scour pool development are primarily driven by sediment impacts on the bed, which are controlled by the spatial distribution of shear stress. We selected the high-flow simulations to examine the distribution of shear stress because it exhibits a pronounced shift of the maximum velocity toward the bed with increasing constriction (velocity inversion), indicating potentially enhanced near-bed acceleration and shear. Introduction of lateral constrictions markedly alters the distribution of bed shear stress under high flow conditions. With no constriction the shear stress is variable upstream of the pool but below 1.0 Pa and in the pool the shear stress is relatively low ($\approx1.0$ Pa) (Figure \ref{Fig7}). Adding a constriction creates elevated shear stress through the constriction at the top of the pool and at the pool entrance, together with a distinct patch of high shear stress near the upstream bottom of the pool. These high-stress regions intensify with increasing constriction and become most pronounced at 35\% constriction (C23), where they also extend farther downstream into the pool. In C23, elevated shear stress persists beyond 2.5 m and spreads to about 2.6 m along the channel. At higher constrictions (C24 and C25), shear stress remains elevated near the constriction and upper pool entrance, but the downstream pattern becomes less patchy and local peak stresses are lower than in C22 and C23. In C24, the main elevated-stress region is concentrated between about 2.3 and 2.5 m, whereas in C25 it is more limited, extending mainly from about 2.3 to 2.4 m. In addition, C25 shows a more distributed pattern of shear stress along the channel bed, rather than the more localized patches observed in C22 and C23.

\vspace{-10pt}

\subsubsection{Visualization and analysis of streamlines}
Streamlines reveal patterns in the flow field that emerge with the introduction of the lateral constriction (Figure \ref{Fig8}). In the absence of lateral constriction, no eddies develop on the surface (Figure \ref{Fig8}a), and the flow remains laterally uniform, with nearly parallel streamlines and minimal velocity variation along the channel. The vertical variation on velocity is also negligible, except for when streamlines diverge as flow enters the pool (Figure \ref{Fig8}b).  When a constriction is introduced, two counter-rotating lateral separation eddies form immediately downstream of the constriction (Figures \ref{Fig8}c). These eddies are bounded by a return current that reconnects with the mainstream flow, creating a well-defined eddy eye characterized by comparatively low velocities (with velocity reversals). The vertical variation in the streamlines also reveals strong downward vertical velocities at the constriction (Figure \ref{Fig8}d), which generate upwelling surface boils near the channel walls immediately downstream of the constriction. This vertical recirculation is consistent with the secondary vortex structure observed by \cite{hunt_experiments_2018}. Both lateral and vertical eddies enhance turbulent mixing and promote velocity inversion, by shifting the high velocities toward the bed. The results further indicate that lateral constriction, rather than the presence of the pool, is the primary driver of eddy formation in this system. 

\begin{figure}[htbp]
    \centering
    \includegraphics[width=\textwidth]{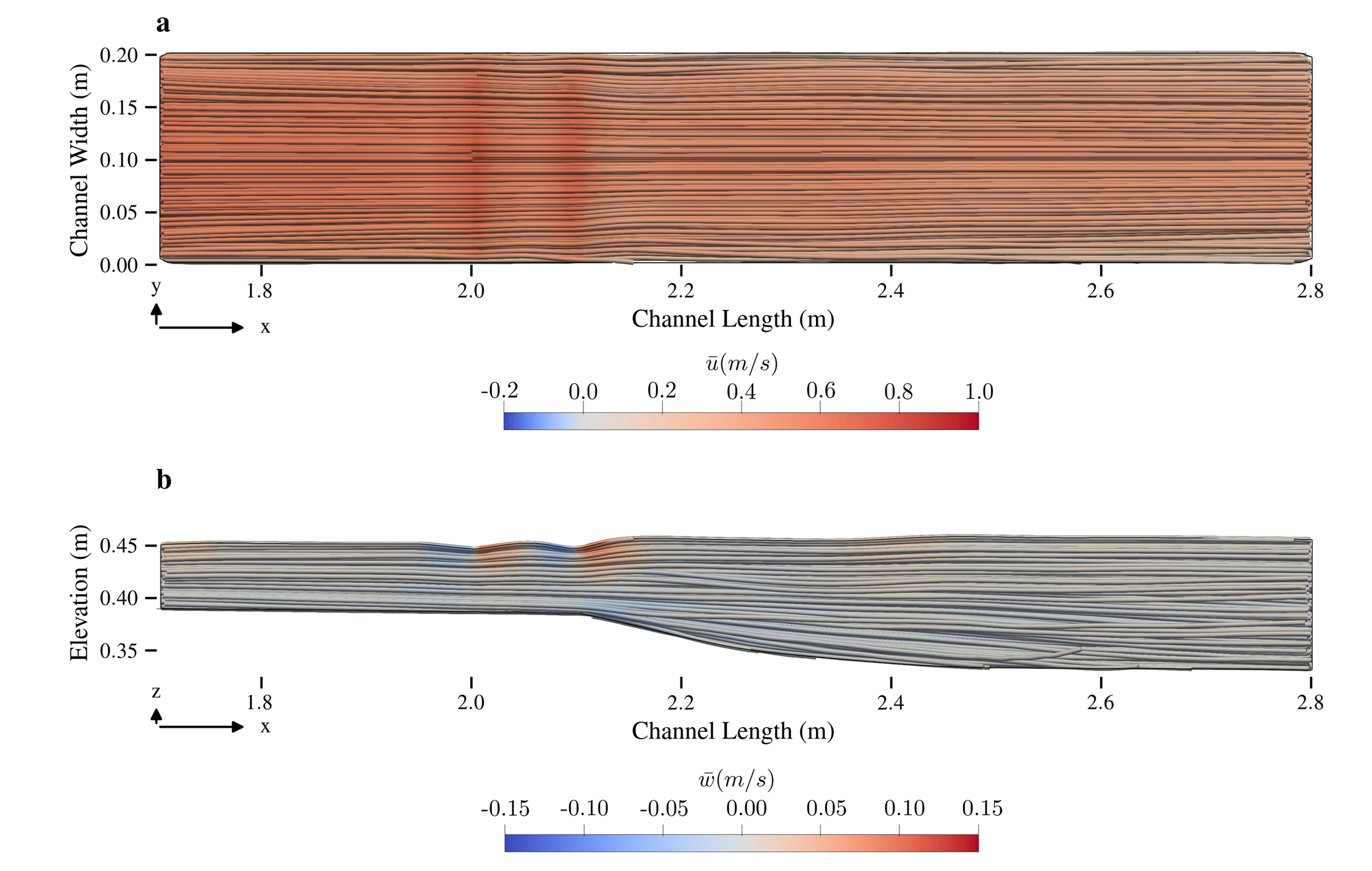}
    
\end{figure}

\clearpage

\begin{figure}[htbp]
    \centering
    \includegraphics[width=\textwidth]{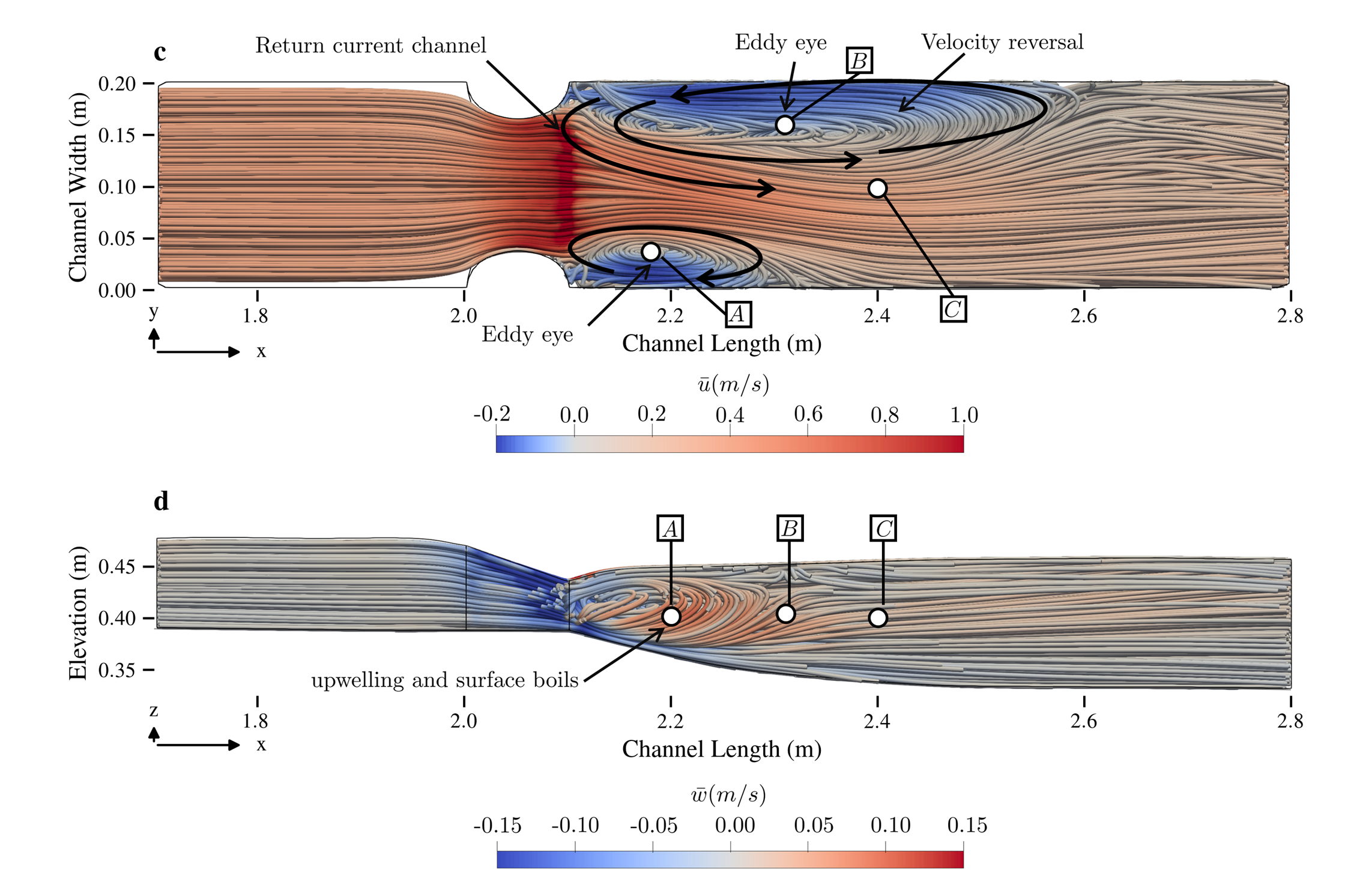}
    \caption{Variation in streamlines with high flow simulations with no constriction (HF0) a) on the surface and b) along a transect through the channel. Streamlines for simulations with a 50\% constriction (C25) c) on the surface and d) along a transect through the channel. Locations A, B, and C are locations in the flow where we examined the oscillation patterns in turbulent flow field.}
    \label{Fig8}
\end{figure}

\subsection{Turbulent flow field}
\subsubsection{Turbulence intensities}
\begin{figure}[htpb]
    \centering
    \includegraphics[width=\textwidth]{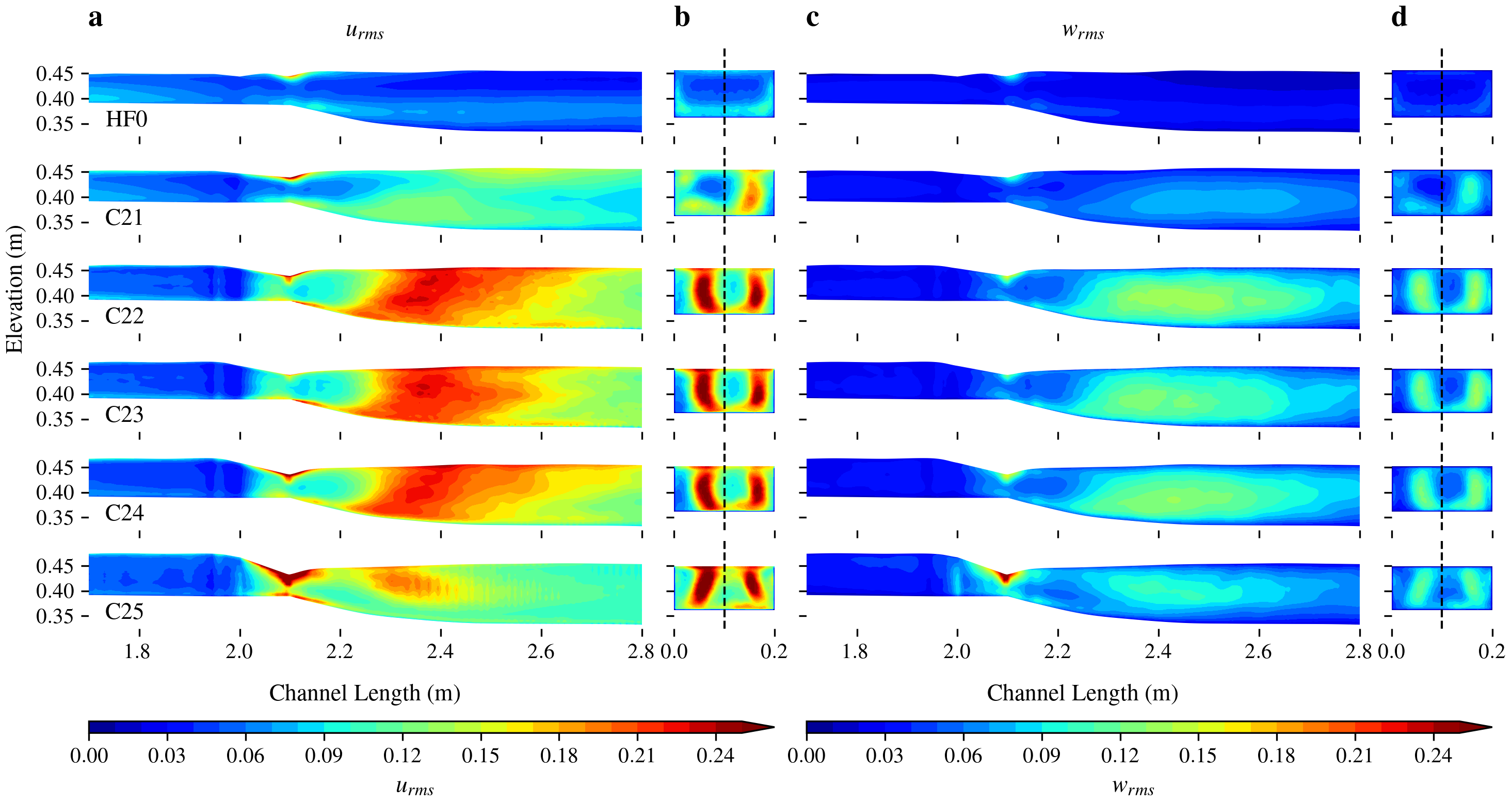}
    \caption{Distributions of a) $u_{rms}$ along the channel centerline, b) $u_{rms}$ for a cross-section at x = 2.2m (c) $w_{rms}$ along the centerline and (d) $w_{rms}$ for a cross-section at x = 2.2m.}
    \label{Fig9}    
\end{figure}

We explored the distribution of turbulence intensity using the root-mean-square velocity fluctuations in the streamwise and vertical directions, $u_{rms}$ and $w_{rms}$ for the high flow case where plunging flows ranged from weak to strong in terms of their development. With no constriction (HF0) present the mean velocity field remains relatively uniform with the highest velocities near the surface and lower velocities near the bed. Turbulence intensity is low throughout the domain, with only slightly elevated values of $u_{rms}$ and $w_{rms}$ near the bed due to local bed shear (Figure \ref{Fig9}a \& c). The cross section at x = 2.2m shows the same uniform behavior in $u_{rms}$ and $w_{rms}$ with elevated values near the boundary (Figure \ref{Fig9}b \& d). \\

\noindent Simulated turbulence intensities vary with the degree of constriction, revealing the turbulence dynamics of plunging flows.  With a 20\% constriction and a weak inversion (C21), the plunging flow results in higher turbulence intensities over the pool (Figure \ref{Fig9}a \& c).  At higher constrictions (C22, C23, C24) with moderate/strong plunging flows, the turbulence intensities are amplified downstream of the constriction (Figure \ref{Fig9}a \& c). Cross-sections of both $u_{rms}$ and $w_{rms}$ reveal higher turbulence intensities at the edges of the plunging flow, suggesting a shear layer forms (Figure \ref{Fig9}b \& d). The core of maximum velocity (CMV) does not coincide with the highest fluctuation intensity. Instead, the core interior exhibits comparatively low $u_{rms}$ and $w_{rms}$, whereas stronger fluctuations develop outside the core. \\

\noindent With a 50\% constriction and strong plunging flow behavior (C25), the turbulence structure changes more substantially. The high-velocity core migrates downward toward the bed, forming a more distinct plunging core, while velocities near the free surface become comparatively lower. These reduced near-surface velocities coincide with elevated $u_{rms}$ and $w_{rms}$, which are clearly visible above the high-velocity core. This behavior differs from C22, C23 and C24, where the accelerated core remains stronger in the mid-to-upper part of the flow. In C25, the CMV is stronger and more closely aligned with the centerline region, and no similarly distinct zones of elevated $u_{rms}$ and $w_{rms}$ are observed within the CMV or in the near-bed region. Overall, the high-velocity CMV exhibits lower velocity fluctuations and shear layers form adjacent the CMV. This indicates that plunging flows enhance turbulence mainly at the margins of the core, where strong velocity gradients promote shear-driven mixing, downwelling, and displacement of the surrounding flow.

\vspace{-10pt}

\subsubsection{LES-Resolved Stresses and Total Turbulent Kinetic Energy}

\begin{figure}[htpb]
    \centering
    \includegraphics[width=\textwidth]{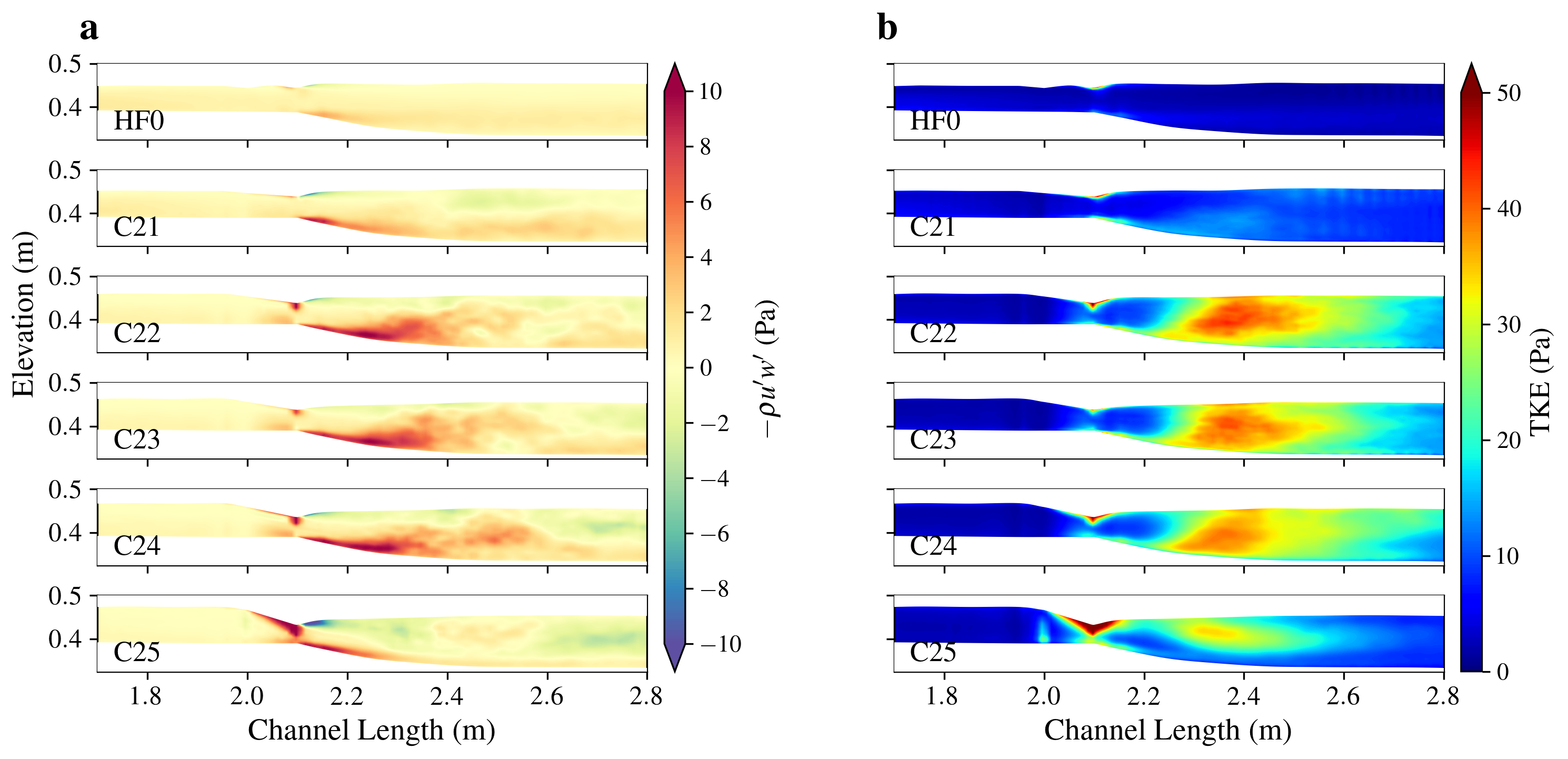}
    \caption{ Spatial distribution of turbulence in plunging flows developed at high flow, including a) the downstream and vertical component of Reynolds stress ($\rho u' w'$) and b) turbulent kinetic energy.}
    \label{Fig10}
\end{figure}

We examine the x – z stress tensor to better understand turbulent shear stress and momentum mixing in plunging flows. Under no-constriction conditions (HF0; Figure \ref{Fig10}a), turbulent mixing remains minimal upstream of the pool, and Reynolds shear stress is weak throughout most of the domain, indicating limited turbulent momentum exchange and the absence of a well-developed plunging structure. The constriction causes shear stress to increase sharply, forming a distinct dip that marks the transition from normal flow with the highest velocity at the surface to a plunging flow. This localized acceleration enhances downward velocity fluctuations while maintaining streamwise alignment of the horizontal velocity components. As the flow enters the pool, strong downward vertical velocities develop along the channel centerline and are directed toward the bed. To balance this motion, upwelling vertical velocities form near the channel walls, displacing lower-velocity fluid toward the surface, while the streamwise flow component remains predominantly aligned with the main flow direction. This dynamic produces positive Reynolds shear stress ($\rho u' w'$) near the bed and negative shear stress near the surface.\\

\noindent The magnitude of Reynolds shear stress response increases with channel constriction, indicating stronger vertical momentum transfer and more energetic, turbulent mixing within the pool. The inflow velocity decreases with increasing channel constriction (Table \ref{Table 2}). Notably, for channel constrictions between 20\% (C21) and 40\% (C24) momentum transfer is dominated by upwelling turbulent mixing by surface boils. In contrast, the 50\% channel constriction (C25) produces a strong plunging flow that results in dominant upwelling turbulent mixing downstream of the constriction. 

\vspace{-10pt}

\noindent Turbulent Kinetic Energy (TKE) represents the energy extracted from the mean flow by the motion of turbulent eddies \parencite{bradshaw_compressible_1977,klein_attempt_2005}. TKE production involves interactions of the Reynolds stresses with mean velocity gradients and, ultimately, TKE dissipation occurs via viscous forces after being passed through the inertial subrange of the turbulence spectrum \parencite{tennekes_first_1972,tennekes1972first}. For the flow with no constriction (HF0), the TKE distribution is generally low throughout the domain, indicating weak extraction of turbulent energy from the mean flow (Figure \ref{Fig10}b). Once a constriction is introduced, TKE increases sharply at the constriction and reaches its maximum immediately downstream at the pool entrance, showing that this is the primary zone where turbulent energy is extracted from the mean flow (Figure \ref{Fig10}b). Among the constricted cases, TKE is greatest for C23 and then decreases with further increases in constriction from C24 to C25. This pattern suggests that moderate constriction (35\%) most effectively converts mean-flow energy into turbulence. At C23, the flow is accelerated strongly enough through the constriction to generate intense shear, plunging, and velocity fluctuations, while still retaining sufficient inflow momentum to sustain high turbulent energy within the pool. Although greater constrictions intensify local acceleration and velocity gradients, the accompanying reduction in inflow velocity decreases the mean-flow energy available for conversion into turbulence. As a result, turbulence remains strong, but the TKE magnitude decreases relative to C23. These results indicate that TKE in the pool reflects a combined effect of constriction-driven local acceleration, plunging-induced shear, and the reduction in inflow velocity with increasing constriction. 

\subsection{Turbulence production (P) \& eddy viscosity  ($\epsilon$)}
\begin{figure}[htbp]
    \centering
    \includegraphics[width=\textwidth]{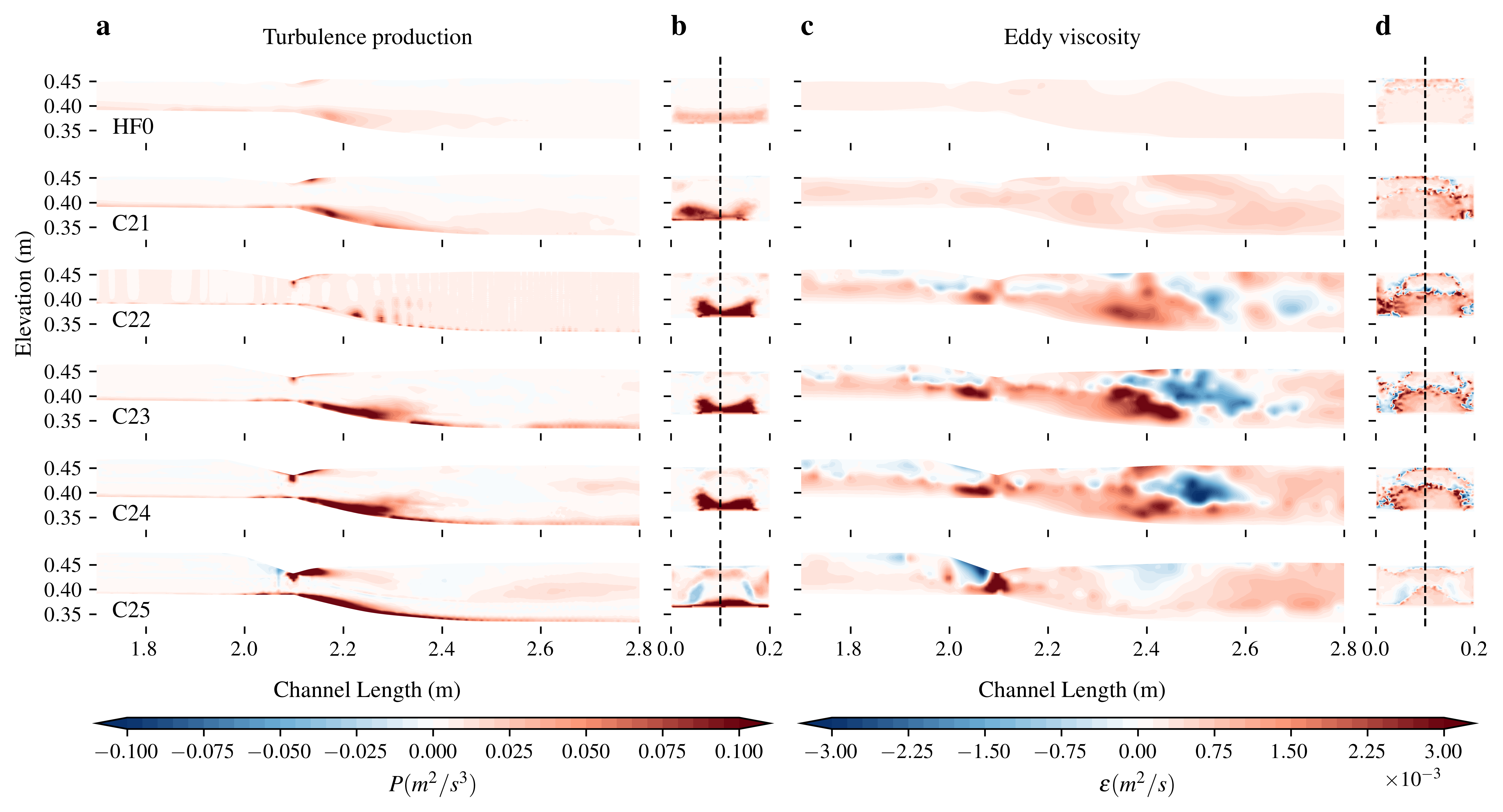}
    \caption{High flow distributions of a) Turbulence production along the channel centerline, b) turbulence production for a cross-section at x = 2.2m (c) eddy viscosity along the centerline and (d) eddy viscosity for a cross-section at x = 2.2m.}
    \label{Fig11}
\end{figure}

We explored the distribution of turbulence production associated with vertical shear for the high-flow case, where plunging flows ranged from weak to strong. With no constriction present (HF0) P is similar to an open-channel flow. Production is mainly concentrated near the bed, where vertical shear is stronger, while lower appears toward the upper part of the flow (Figure \ref{Fig11}a). The cross-section at x = 2.2m shows a similar structure, with positive P mainly located in the lower flow region (Figure \ref{Fig11}b).\\

\noindent Turbulence production varies more strongly with the channel constriction. With a 20\% constriction and weak inversion (C21), P  remains mainly concentrated near the bed, indicating that the velocity inversion is not yet fully developed and that turbulence generation is still dominated by near-bed shear (Figure \ref{Fig11}a). At higher constrictions (C22, C23, and C24), where moderate plunging-flow behavior develops, higher P is consistently observed as the flow enters the pool immediately downstream of the constriction. Farther downstream, the positive P layer becomes thinner and its magnitude decreases. This suggests that vertical shear weakens as the plunging flow develops and the momentum field becomes more redistributed within the pool. The cross-sections show the same trend, where the positive P layer becomes progressively elevated and thicker from C21 to C24 (Figure \ref{Fig11}b). The core of maximum velocity generally coincides with positive P.\\

\noindent With a 50\% constriction and strong plunging-flow behavior (C25), the P structure changes more substantially. Alongstream, C25 develops a distinct layered pattern, with strong positive P near the bed, followed by a negative layer, then a weaker positive layer, and finally a weak negative region higher in the flow (Figure \ref{Fig11}a). This alternating positive–negative structure indicates that the strong constriction reorganizes the vertical shear field and creates multiple zones of energy exchange between the mean flow and turbulent fluctuations. At the x = 2.2m cross-section, the positive P layer becomes thinner and more sharply concentrated compared with C22, C23, and C24 (Figure \ref{Fig11}b). The core of maximum velocity remains associated with positive P, while distinct negative patches are visible outside the core, especially in C25.\\

\noindent Distributions of eddy viscosity for the high flow simulations show how turbulent momentum exchange changes with increasing channel constriction (Figure \ref{Fig11}c–d). In HF0, $\epsilon$ remains relatively low and spatially uniform, indicating limited turbulent momentum mixing in the absence of constriction. With 20\% constriction (C21), $\epsilon$ slightly increases as the flow enters the pool, suggesting that the developing plunging flow begins to enhance momentum exchange. For the intermediate constriction cases (C22, C23, and C24), $\epsilon$ becomes stronger and more spatially heterogeneous within the pool, with positive regions extending downstream of the constriction.\\

\noindent However, C25 shows a different behavior. Although flow velocities decrease with increasing constriction, $\epsilon$ in the pool is lower in C25 compared with C22–C24. This suggests that the strongest constriction does not promote broadly distributed turbulent momentum mixing throughout the pool. Instead, C25 appears to form a more coherent and concentrated plunging-flow core, where momentum exchange is localized rather than widely spread. At the cross-section (Figure \ref{Fig11}d), positive $\epsilon$ is observed near the plunging-flow core, while negative regions appear around the outside of the core. These negative patches are most distinct in C25 and indicate localized counter-gradient momentum transport or reduced mixing efficiency outside the main high-velocity pathway.

\subsubsection{Counter rotating eddies and flow oscillations}

\begin{figure}[h!]
    \centering
    \includegraphics[width=\textwidth]{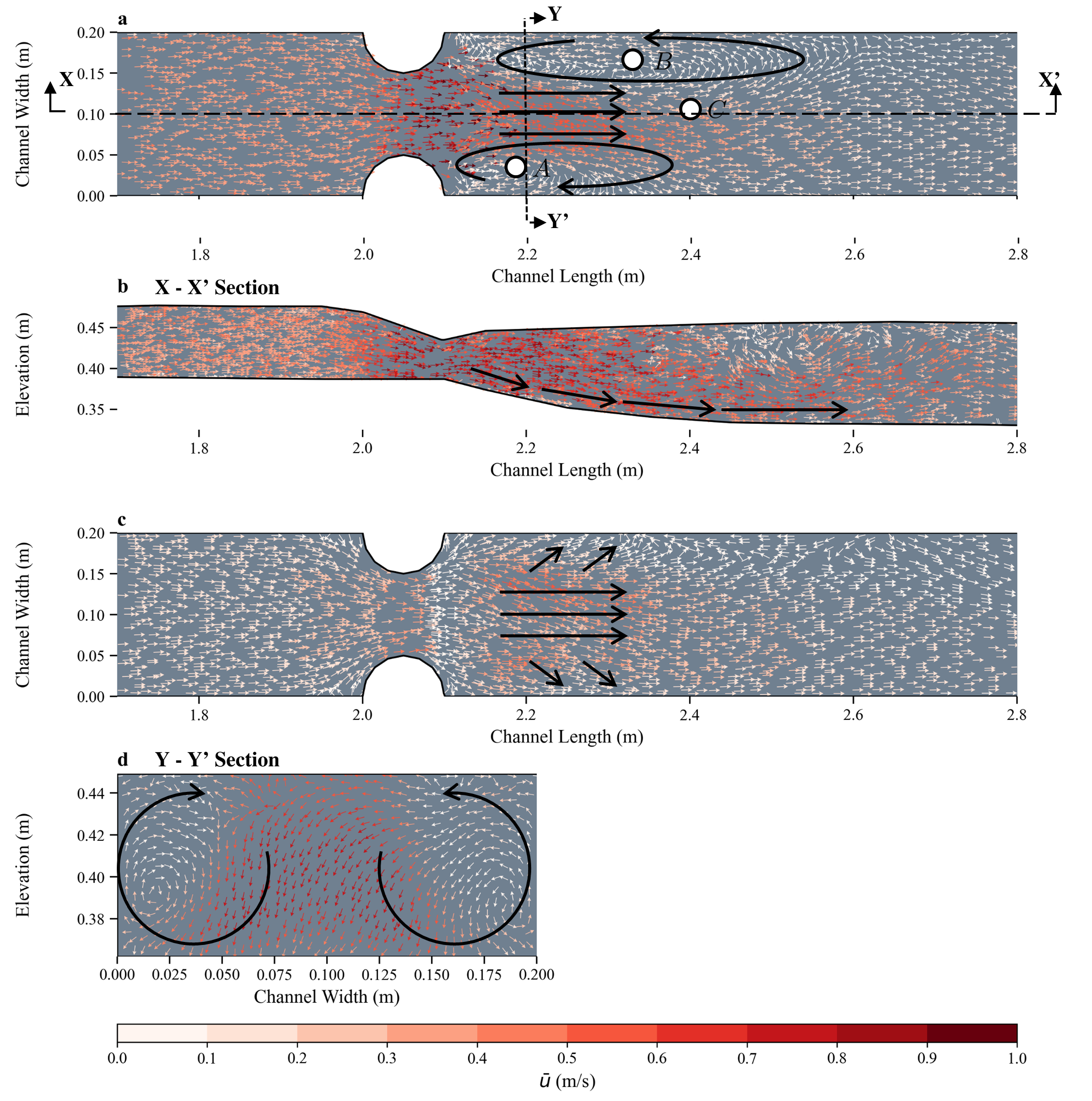}
    \caption{Velocity vectors for high flow with a 50\% constriction (C25) along (a) at the water surface   (b) an along stream transect through the plunging flow, (c) 2mm above the bed, and d) through a cross-section downstream of the plunging flow.}
    \label{Fig12}
\end{figure}

We explore the dynamics of turbulent eddies in the flow using velocity vectors fields. We chose to examine the high flow scenario with the 50\% constriction because it has the best developed plunging flow. At the water surface, flow is accelerated through the constriction, followed by convergent decelerated flow into the pools, and intermittent surface boils generated by large-scale coherent flow structures downstream of the constriction (Figure \ref{Fig12}a). There are asymmetric counter rotating eddies that form immediately downstream of the constriction widening. The asymmetry oscillates back and forth across the channel. Moreover, velocity vector field sliced through the flow field in the x-z plane shows a plunging flow core, with higher-velocity zones forming near the bed immediately after the lateral constriction and progressively shifting deeper downstream (Figure \ref{Fig12}b). At the bed divergent accelerated flow occurs through the constriction, followed by decelerated flow into the widening, with zones of upwelling that exhibit large-scale coherent flow structures (Figure \ref{Fig12}c). A cross-section through the plunging flow reveals downwelling in the channel center and upwelling along the bank that forms surface boils (Figure \ref{Fig12}d). \\

\begin{figure}[htbp]
    \centering
    \includegraphics[width=\textwidth]{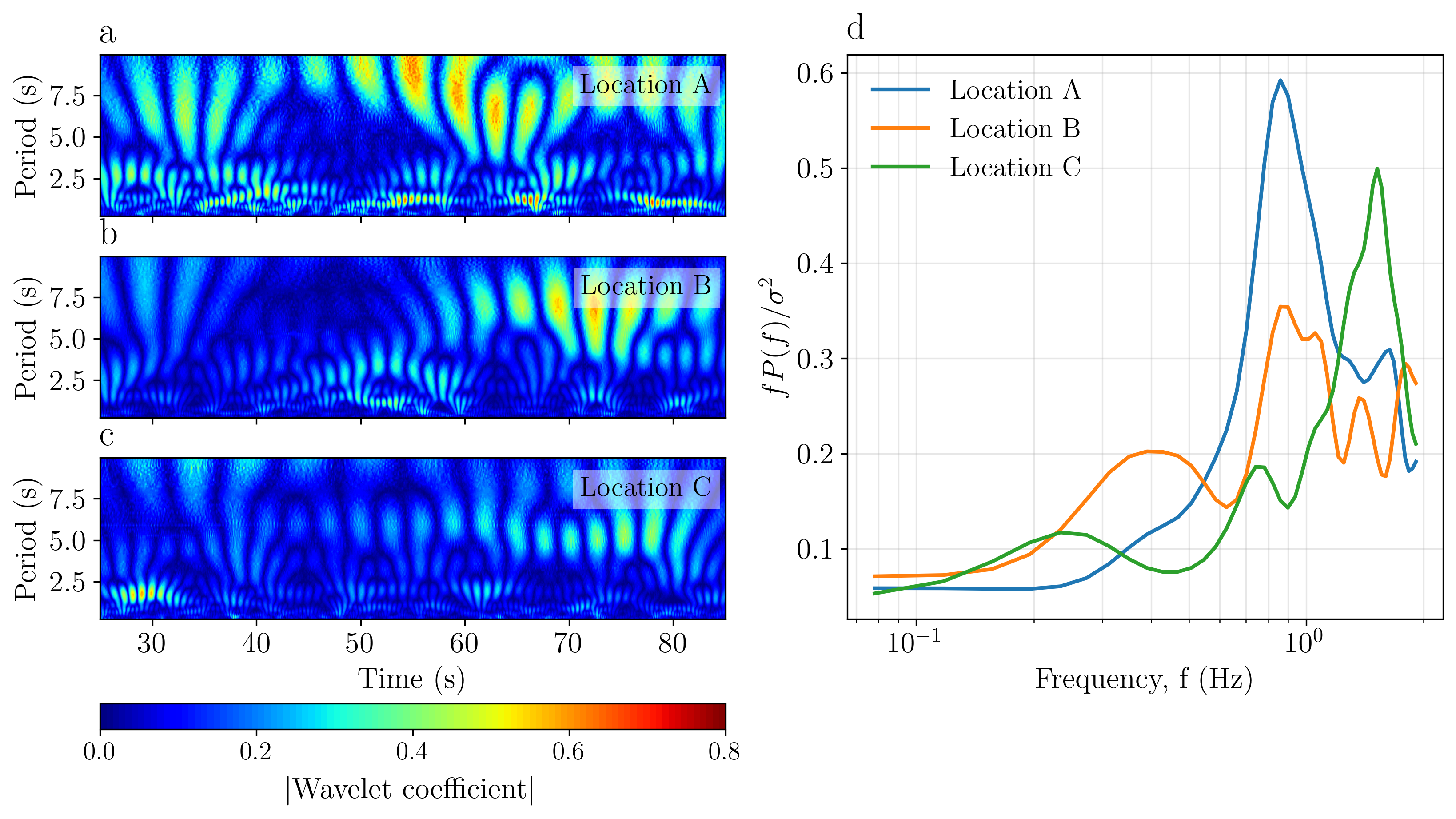}
    \caption{Continuous wavelet analysis of velocity-magnitude oscillations for the 50\% constriction case (C25) at Location A (a), Location B (b), and Location C (c). Panel (d) shows the normalized pre-multiplied spectrum, $fP(f)/\sigma^2$, as a function of frequency, representing the variance-preserving form of the spectral density. }
    \label{Fig13}
\end{figure}

\newpage
\noindent To examine this oscillatory behavior in the counter rotating eddies and its effect on plunging flow, we selected three monitoring locations: Locations A and B correspond to the recirculation eddy cores while location C corresponds to the CMV (Figure \ref{Fig8}). We extracted velocity-magnitude time series at approximately 20\% flow depth below the water surface. The velocity was sampled at 0.1s intervals for 60s, from 25 to 85 s of simulation time to exclude the initial model spin-up period. We then applied continuous wavelet transform analysis to characterize the temporal oscillations and their dominant periodic structures (Figure \ref{Fig13}a–c). The scalograms showed multiple dominant periods of oscillation that appear in patches, indicating intermittent oscillatory behavior, making it difficult to identify a single dominant oscillation scale. Therefore, we further analyzed the velocity spectra in variance preserving form to identify the dominant oscillation frequencies and associated energetic eddy scales.

\noindent For the C25 case, the variance-preserving spectra shows clear spatial differences in the dominant velocity-magnitude oscillations among Locations A, B, and C. In the recirculation eddies (A and B), the spectrum exhibits a strong and relatively sharp peak near 0.85Hz,
corresponding to a dominant oscillation period of approximately 1.1s. At Location B, the spectrum shows a broader response with energy is distributed across approximately 0.4 – 1.2Hz, but with a peak at the same Frequency as Location A.  At Location C, the dominant peak shifts to a higher frequency near 1.4Hz, corresponding to a shorter oscillation period of approximately 0.7s. This indicates that the plunging flow pulses.  Near the constriction (Location C), the horizontal recirculation eddy does not vary with the CMV pulsation.  Further from the constriction, the pulsation of the CMV syncs with the variation in the horizontal recirculation eddies.

\section{Discussion}
We used large eddy simulations (LES) to investigate plunging flows in bedrock rivers, which are difficult to study through field and laboratory experiments due to practical constraints.  Our central goals were to see how well an LES model could represent laboratory-scale plunging flows and, assuming the flows could be accurately simulated, to examine their dynamics at a resolution not possible in the field.    

\subsection{Can subcritical plunging flows be reproduced in a numerical model?}
\noindent The LES model was able to simulate the laboratory scale flows well, particularly under conditions where plunging flow was strongly developed. For the high flow cases, the model reproduced the same overall pattern observed in the laboratory experiments, with plunging-flow strength increasing as discharge increased. This agreement indicates that, when the hydraulic conditions are favorable for plunging flow development, the simulations can capture the dominant flow structure and are associated with plunging flows very well. However, the model performance was less consistent under low-flow conditions. In the laboratory experiments, weak plunging flows were observed at low discharge, but the simulations did not consistently produce the same plunging-flow behavior. This suggests that the model is less reliable under marginal or transitional conditions where plunging flow is weak. Therefore, although the simulations reproduce the main discharge-dependent trend in plunging-flow strength, the low-flow cases should be interpreted with caution. The high-flow scenarios provide the strongest basis for understanding the dynamics of plunging flow, which is why our main interpretation focuses on cases where plunging flow is well developed, and model performance is strongest.\\

\noindent The \cite{hunt_experiments_2018} laboratory experiments showed that the strongest controls on plunging flows were the amount of the channel constricted and the Froude number.  There was an increase in plunging flow intensity with constriction percentage, but that resulted in an increase in the Froude number and the strongest plunging flows occurred in transcritical flows (0.8 < Fr < 1.2).  Yet field observations of plunging flows have been universally subcritical with Froude numbers rarely exceeding 0.5 \parencite{wright_covariation_2022}.  For example, the Fraser River that has hundreds of documented plunging flows \parencite{hurson_abundance_2025}, <4.5\% of the canyon distance has Fr>0.5; <0.5\% has Fr>0.6 and <0.05\% has Fr>0.7.  This would suggest either that the flows produced at laboratory scale are produced by a fundamentally different process, akin to a submerged hydraulic jump than those produced at field scale, or that there is something suppressing the Froude number at field scale, like high bank roughness that reduced the velocity. \\

\noindent Our simulations indicated that plunging flows remained subcritical at the constriction, with an amplified Froude number approaching but never exceeding the critical stage. Upstream and downstream of the constriction, the flow consistently remained in the subcritical regime. Our simulations supported the positive correlation between channel constriction and the Froude number. Both high and low flow exhibited Froude number amplification with velocity inversion strength.  That our simulations remained subcritical through the constriction is encouraging because it would suggest that we could simulate field scale flow using the LES techniques developed here without introducing a Froude number damping effect for the field.  Nevertheless, the laboratory scale flow and the simulations produced the best plunging flows for transcritical conditions.  The discrepancy in Fr between field and laboratory scale remains an unsolved problem.

\subsection{What are the spatial and temporal characteristics of plunging flow structures?}
\noindent The overall pattern of simulated velocity in the simulations support the conceptual model from \cite{venditti_flow_2014} who described plunging flows as a centreline velocity inversion that is coincident with counter rotating secondary circulation driven by upwelling coherent flow structures along the banks. They described the main cause as lateral constriction, which increases the water surface upstream of the constriction.  Interpolating from their description of plunging flows, it is expected that plunging flows would get stronger as discharge increased and as channel constriction increased.  This was later supported by experiments \parencite{hunt_experiments_2018} and field observations \parencite{hurson_amplification_2022}. Our simulations also support discharge and constriction controls on plunging flows, but with some caveats.  Greater constriction does not always cause increased velocity inversion.  The simulations show that constriction does increase the strength of plunging flow at high flows with the most well-defined velocity inversion occurring 50\% constriction.  But at low flow, the best-defined plunging flow happens at 35\% constriction.  The best-defined plunging flow, in terms of velocity inversion, is not the one that has the highest shear stress.  The Reynolds stress field at the high flow shows the highest turbulent stresses and greatest bed stresses (from velocity gradients) occur at 35\% constriction.  This has important implications for understanding plunging flows and suggests that there are optimal combinations of discharge and channel constriction that produce different effects.\\

\noindent The reason for these patterns is linked to the nature of turbulence in the non-uniform flow fields.  In the absence of channel constriction, turbulence levels, production, and mixing are primarily controlled by bed-induced shear.  Turbulent intensity distributions show that constricted plunging flows enhanced turbulence, mainly at the margins of the core, where strong velocity gradients promote shear-driven mixing, downwelling, and displacement of the surrounding flow occur.  Higher turbulence production is consistently observed as the flow enters the pool immediately downstream of the constriction. This indicates that the strongest shear-driven energy transfer occurs near the pool entrance, where the constricted flow adjusts rapidly to the expanded pool geometry. Farther downstream, the high turbulence production layer becomes thinner and its magnitude decreases. This suggests that vertical shear weakens as the plunging flow develops and the momentum field becomes more redistributed within the pool.  The turbulence-production distributions show that plunging flows enhance turbulence generation mainly through shear redistribution around the CMV while stronger constriction produces more distinct positive and negative production layers associated with velocity inversion and localized counter-gradient momentum exchange.\\

\noindent Constrictions intensify momentum exchange between the CMV, the upper flow, and the near-bed region, as indicated by Reynolds stress distributions. Enhanced turbulent mixing and redistribution of momentum occurs within the pool for constricted flows. TKE increases sharply at the constriction and reaches its maximum immediately downstream at the pool entrance, showing that this is the primary zone where turbulent energy is extracted from the mean flow.  The eddy viscosity distributions show that moderate constrictions enhance spatially distributed turbulent mixing, whereas the greatest constriction organizes momentum exchange into a narrower region around the plunging-flow core.  This indicates that moderate constrictions promote broader turbulent mixing as the plunging flow interacts with the surrounding slower-moving water.\\

\noindent Our simulations reveal that plunging flows are intermittent, getting stronger and weaker through time.  That suggests some of the variability we observe in the field \parencite{hurson_abundance_2025}, may be due to the intermittency of plunging flows. Similar behavior was observed in the velocity time series from the simulations. The flow did not oscillate at one constant rate. Instead, energetic motions appeared and disappeared through time. This intermittency also varied spatially across the flow field. The recirculation zones showed similar dominant frequencies, suggesting that these regions fluctuated in a related way. However, the CMV showed faster, shorter-term oscillations. This indicates that the plunging flow pulses through time, but that the timing and strength of these pulses vary across the flow field.

\noindent These patterns of turbulence impact the distribution of shear stress on the bed and therefore have important implications for how plunging flows erode bedrock.  Shear stress calculated from the velocity gradient shows that although inlet velocity decreases with increasing constriction, the constriction reduces the cross-sectional area and locally accelerates the flow, producing high velocity gradients and elevated shear stresses at the constriction. The constriction also modifies the downstream flow structure by inducing velocity inversions, while a backwater effect increases upstream water depth and amplifies flow through the constriction. Together, these processes enhance near-bed velocity gradients and elevate shear stress within the pool. With no constriction, our simulations show that the maximum velocity remains relatively high in the water column, so near-bed velocities and shear stresses in the pool remain low. For constrictions ranging from 20\% to 35\%, the high-velocity core shifts progressively downward and penetrates farther into the pool, strengthening near-bed velocity gradients and producing increasingly strong, patchy bed shear stresses. This effect is most pronounced for 35\% constriction at high flow, which retains a slightly higher velocity than at constrictions of 40\% and 50\%, allowing stronger localized near-bed gradients and correspondingly higher localized shear-stress patches (Figure 5). In for high levels of constriction, (40\% and 50\%), the maximum velocity remains low and elevated velocity extends farther downstream, but the high-velocity region becomes more broadly distributed, resulting in a wider and less patchy shear-stress pattern. Overall, the shear-stress distributions are consistent with the velocity, indicating that bed shear stress is controlled by local flow acceleration, the vertical position of the high-velocity core, its spatial concentration, and inlet velocity.\\

\subsection{Do numerical simulations support or challenge existing hypotheses on plunging flow formation?}
\noindent The first observations of plunging flows in bedrock canyons described them as occurring due to convective acceleration through a constriction and deceleration into a pool that creates a 3D secondary flow structure that suppresses the CMV below the water surface \parencite{venditti_flow_2014}. Subsequent investigations have indicated Froude number is an important control, while distinguishing them from submerged hydraulic jumps \parencite{hunt_experiments_2018}.  Recent work has emphasized the importance of backwater formation at a constriction that produces a spilling flow through the constriction and the velocity inversion \parencite{cao_experiments_2022,kusack_experiments_2024}. These explanations are not mutually exclusive but rather are different aspects of the process that leads to plunging flows.\\

\noindent Our simulations broadly support these explanations.  We do observe convective acceleration and deceleration through a constriction and the formation of counter rotating, vertical secondary flow cells downstream of a constriction.  Field observations emphasize upwelling of low-velocity fluid toward the surface in the form of large-scale coherent flows structures (kolks) with a boils forming at the water surface.  The flow is envisioned to fold over on itself with the low speed fluid rising up the walls and pushing the CMV towards the bed as it breaches the surface \parencite{venditti_flow_2014}. Our simulations showed that there are persistent horizontal recirculation cells downstream of constrictions and the formation of kolks and boils are not as obvious.  This overall pattern is something that can be searched for in future field observation.\\

\noindent No distinct backwater effect was observed in the simulations. This is not surprising because the LES simulations employed a rigid-lid approach based on the observed water surface, which limited the possible expression of a strong backwater to what was observed in the experiments.  Under low-flow conditions with maximum width constriction, velocity reversals indicative of backwater formation, were detected. This effect was absent at higher discharges, but the effect of the rigid-lid is more important at higher flows.  Nonetheless, the influence of backwater was not sufficiently pronounced to be clearly distinguished in the simulations.
Our simulations lead to a more refined explanation of why plunging flows occur.  After passing through a constriction, the flow re-expanded and undergoes convective deceleration under subcritical conditions (Fr < 1). The associated adverse pressure gradient slows the upper layers while redirecting momentum downward and outward, concentrating higher velocities near the bed and sidewalls. This behavior reflects momentum redistribution, where high-momentum fluid from the core is displaced toward the lower region as the flow adjusts to the wider downstream section.  Concurrently, partial pressure recovery occurs as part of the kinetic energy is converted back into static pressure, while some energy dissipates through turbulence and mixing. Together, these mechanisms explain the observed near-bed velocity inversion and gradual backwater effect characteristic of subcritical plunging flows. This leads to an increase in potential energy upstream of the lateral constriction. As a result, the flow must move vertically downward to reach equilibrium. The elevation difference between the upstream and downstream of the lateral constriction leads to the formation of high-velocity cores and velocity inversions. The reduction in cross-sectional area at the lateral constriction results in high velocities contributing to the acceleration of water flow.\\

\noindent While our modelling broadly supports previous explanations for why plunging flows occur, more work is necessary to understand how plunging flows change, particularly in the field.  \cite{hurson_amplification_2022} observed amplification of plunging flows as discharge increases, resulting ~15 meters of sediment cover evacuation.  Our simulations support this pattern of plunging flows getting stronger as discharge increases.  However, the insights into how plunging flows change in the field, laboratory, and our numerical model are all limited to comparison of just two flows.  \cite{hurson_amplification_2022} was only able to compare plunging flows at low and moderate discharges because it is not logistically possible to measure plunging flows at high flows in bedrock canyons.  Our simulations and experiments focused on two flows because it is a tractable problem.  What happens to plunging flows at field scale during high discharge is mostly unknown.  Future work needs to adapt the numerical simulation techniques developed here for laboratory scale and apply them to the field scale.  There is a need simulate plunging flow dynamics at a range of discharges that include those where the is validation data at low flows \parencite{hurson_amplification_2022,hurson_abundance_2025,venditti_struggles_2020,venditti_flow_2014} and moderate flows \parencite{hurson_amplification_2022}.  Critically, field scale simulations are needed for flows above those that can be measured in the field, ranging from moderate flows to the highest flows that might reasonably be expected in a bedrock canyon.  These high flow simulations are necessary to understand patterns of shear stress, sediment transport, and bedrock erosion, which would allow modelling of canyon evolution that includes our understanding of plunging flow dynamics.\\

\section{Conclusions}

This study provides the first ever high-resolution modeling of the three-dimensional turbulent flow structure of plunging flows in bedrock rivers, offering new insights into their structure, controls, and implications for bed erosion. From our results we can conclude:

\begin{enumerate}
    \item Plunging flows respond to increased discharge with enhanced turbulence-driven momentum exchange. Downstream of the constriction, the development of plunging flow corresponds with regions of high energy dissipation, with turbulent kinetic energy peaking at the pool entrance immediately downstream of the constriction. Surface upwelling (counter-rotating surface boils) and divergent near-bed flows further supported the presence of convoluted, 3D convective structures.
    \item Channel constriction plays a critical role in amplifying turbulence by generating strong shear layers and inducing velocity profile inversions. These inversions led to the formation of high-shear zones near the bed and along the constriction, which were absent in unconstricted scenarios.
    \item Both discharge and constriction show strong positive controls on plunging flow intensity, but their relationship is non-linear, with optimal combinations of the two producing distinct effects on flow structure and erosive potential.
    \item Plunging flows can form under subcritical conditions through channel constrictions, but the best defined plunging flows happen under transcritical conditions. Both discharge and constriction are positively correlated with Froude number amplification and velocity inversion strength. There is a discrepancy between universally subcritical plunging flows at field scale and transcritical plunging flows at small scales that remains unresolved.
    \item Channel constriction fundamentally reorganizes bed shear stress distributions in plunging flows Moderate constrictions produce strong, spatially concentrated patches of elevated bed shear stress, while greater constrictions distribute stress more broadly but at lower intensity.
    \item Plunging flows are inherently intermittent, strengthening and weakening through time. 
\end{enumerate}

\noindent Future numerical modelling work should explore field scale plunging flow dynamics at discharges too high to make observations in real bedrock canyons.  That would open the possibility to predict stress fields, sediment transport dynamics, and bedrock erosion patterns through constriction-pool-widening channel morphologies.

\section*{Acknowledgments}
This research was supported by the National Science Foundation (NSF) CAREER Award No. 2239550, the National Oceanic and Atmospheric Administration (NOAA) Cooperative Science Center Educational Partnership Program with Minority Serving Institutions (MSI) Award No. NA22SEC4810016, and the Army Research Office (ARO) Cooperative Agreement No. W911NF-23-2-0046 to LVA. an NSF Implementation Grant, Community-driven Inclusive Excellence and Leadership Opportunities in the Geosciences (CIELO-G), Award No. 2228180, supporting JTS; a British Columbia Salmon Restoration Innovation Fund (BCSRIF) grant to JGV and 9 others; as well as a Natural Sciences and Engineering Research Council of Canada (NSERC) Discovery Grant and Accelerator Supplement to JGV.\\
\noindent The views and conclusions contained in this document are those of the authors and should not be interpreted as representing the official policies, either expressed or implied, of ARO, NSF, NOAA, or the U.S. Government.

\section*{Open Research}
The laboratory data used for comparison with the numerical simulations in this study were obtained from Hunt et al. (2018) and are available within the associated publication.
The processed numerical simulation data supporting Figures 1–13 are publicly available through Microsoft OneDrive with read-only access (Samarasinghe et al., 2026). The archive contains the processed figure data, computational-domain geometry files, and streamline visualization files used in the analyses presented in this study.\\

\noindent OpenFOAM v2112, used to conduct the large-eddy simulations, is openly available from OpenCFD Ltd. under the GNU General Public License (OpenCFD Ltd., 2021).\\

\noindent Hunt, B., Venditti, J. G., \& Kwoll, E. (2018). Experiments on the morphological controls of velocity inversions in bedrock canyons. Earth Surface Processes and Landforms, 43(3), 654–668. \url{https://doi.org/10.1002/esp.4274}\\

\noindent Samarasinghe, J. T., Alvarez, L. V., Hurson, M., \& Venditti, J. G. (2026). Data supporting “Large eddy simulation of plunging flows in laboratory-scale bedrock rivers” [Dataset]. Microsoft OneDrive. \href{https://minersutep-my.sharepoint.com/personal/jthambanged_miners_utep_edu/_layouts/15/onedrive.aspx?id=%2Fpersonal%2Fjthambanged%5Fminers%5Futep%5Fedu%2FDocuments%2FSamarasinghe%5Fet%5Fal%5F2026%5FJGR%5FEarth%5FSurface%5FData&ga=1}{Microsoft OneDrive}\\

\noindent OpenCFD Ltd. (2021). OpenFOAM v2112 [Software]. OpenCFD Ltd. \url{https://www.openfoam.com/news/main-news/openfoam-v2112}

\section*{Author Contributions}
Conceptualization, JTS, LVA and JGV; methodology, JTS., LVA and JGV; software, JTS.; formal analysis, JTS and LVA with contributions from JGV; resources, JTS and LVA; data curation, JTS and LVA; writing original draft preparation, JTS, JGV and LVA; review and editing, JTS, JGV, LVA, and MH; visualization, JTS; supervision, LVA and JGV; project administration LVA and JGV; funding acquisition, JTS, LVA, JGV All authors have read and agreed to the published version of the manuscript.

\section*{Variable Index}

\begin{longtable}{
    >{\raggedright\arraybackslash}p{0.18\textwidth}
    >{\raggedright\arraybackslash}p{0.57\textwidth}
    >{\centering\arraybackslash}p{0.17\textwidth}
}

\toprule
\textbf{Variable} & \textbf{Definition} & \textbf{Units} \\
\midrule
\endfirsthead

\multicolumn{3}{l}{\textit{Variable Index continued}}\\
\toprule
\textbf{Variable} & \textbf{Definition} & \textbf{Units} \\
\midrule
\endhead

\midrule
\multicolumn{3}{r}{\textit{Continued on next page}}\\
\endfoot

\bottomrule
\endlastfoot

$A_f$
& Surface area of the grid cell face
& m$^2$ \\

$C_k$
& SGS model constant
& Dimensionless \\

$f$
& Frequency
& s$^{-1}$ \\

$Fr$
& Froude number
& Dimensionless \\

$g$
& Gravitational acceleration
& m\,s$^{-2}$ \\

$h$
& Local flow depth
& m \\

$h_{\mathrm{max\,Vel}}$
& Vertical distance from the bed to the point of maximum velocity
& m \\

$k_{\mathrm{res}}$
& Resolved turbulent kinetic energy
& m$^2$\,s$^{-2}$ \\

$k_{\mathrm{sgs}}$
& Subgrid-scale turbulent kinetic energy
& m$^2$\,s$^{-2}$ \\

$k_{\mathrm{total}}$
& Total turbulent kinetic energy
& m$^2$\,s$^{-2}$ \\

$k_s$
& Equivalent roughness height
& m \\

$\hat{\mathbf{n}}_f$
& Unit normal vector to the cell face
& Dimensionless \\

$\overline{p}$
& Filtered kinematic pressure
& m$^2$\,s$^{-2}$ \\

$P$
& Turbulence production due to vertical shear
& m$^2$\,s$^{-3}$ \\

$P(f)$
& Power spectral density of the detrended velocity signal
& m$^2$\,s$^{-1}$ \\

$S_{ij}$
& Resolved strain-rate tensor
& s$^{-1}$ \\

$t$
& Time
& s \\

$u$
& Instantaneous velocity in the streamwise $x$-direction
& m\,s$^{-1}$ \\

$\overline{u}$
& Mean streamwise velocity
& m\,s$^{-1}$ \\

$u'$
& Instantaneous streamwise velocity fluctuation
& m\,s$^{-1}$ \\

$\overline{u}_i$
& Filtered velocity component in the $i$-th direction
& m\,s$^{-1}$ \\

$\overline{u}_j$
& Filtered velocity component in the $j$-th direction
& m\,s$^{-1}$ \\

$\overline{u}_i\overline{u}_j$
& Product of the filtered velocity components in the $i$- and $j$-directions
& m$^2$\,s$^{-2}$ \\

$\overline{u_i u_j}$
& Filtered product of velocity components $u_i$ and $u_j$
& m$^2$\,s$^{-2}$ \\

$u_{\mathrm{rms}}$
& Root-mean-square streamwise velocity fluctuation
& m\,s$^{-1}$ \\

${u'}^2$
& Instantaneous squared streamwise velocity fluctuation
& m$^2$\,s$^{-2}$ \\

$u'v'$
& Instantaneous product of streamwise and vertical velocity fluctuations
& m$^2$\,s$^{-2}$ \\

$\overline{u'w'}$
& Time-averaged Reynolds shear stress component associated with streamwise and lateral velocity fluctuations
& m$^2$\,s$^{-2}$ \\

$\displaystyle \frac{\partial \overline{u}_i}{\partial x_i}$
& Divergence of the filtered velocity field
& s$^{-1}$ \\

$\displaystyle \frac{\partial \overline{u}_i}{\partial t}$
& Local acceleration of the filtered velocity component
& m\,s$^{-2}$ \\

$\displaystyle
\frac{\partial\left(\overline{u}_i\overline{u}_j\right)}
{\partial x_j}$
& Convective acceleration term
& m\,s$^{-2}$ \\

$\displaystyle \frac{\partial \overline{p}}{\partial x_i}$
& Pressure-gradient term
& m\,s$^{-2}$ \\

$\displaystyle \frac{\partial \overline{u}_i}{\partial x_j}$
& Velocity-gradient tensor component
& s$^{-1}$ \\

$\displaystyle \frac{\partial \overline{u}_j}{\partial x_i}$
& Transposed velocity-gradient tensor component
& s$^{-1}$ \\

$\displaystyle \frac{\partial u}{\partial z}$
& Velocity gradient used in the shear-stress calculation
& s$^{-1}$ \\

$U$
& Velocity magnitude or characteristic velocity
& m\,s$^{-1}$ \\

$U_f$
& Velocity magnitude at the cell face
& m\,s$^{-1}$ \\

$U_{\mathrm{in}}$
& Initial or inlet velocity used in the experiment/model
& m\,s$^{-1}$ \\

$U_{\max}$
& Maximum velocity
& m\,s$^{-1}$ \\

$U_{\mathrm{surf}}$
& Near-surface velocity
& m\,s$^{-1}$ \\

$v$
& Instantaneous velocity in the vertical $y$-direction
& m\,s$^{-1}$ \\

$\overline{v}$
& Mean vertical velocity
& m\,s$^{-1}$ \\

$v'$
& Instantaneous vertical velocity fluctuation
& m\,s$^{-1}$ \\

$v_{\mathrm{rms}}$
& Root-mean-square vertical velocity fluctuation
& m\,s$^{-1}$ \\

${v'}^2$
& Instantaneous squared vertical velocity fluctuation
& m$^2$\,s$^{-2}$ \\

$v'w'$
& Instantaneous product of vertical and lateral velocity fluctuations
& m$^2$\,s$^{-2}$ \\

$V_p$
& Volume of the computational cell
& m$^3$ \\

$w$
& Instantaneous velocity in the lateral $z$-direction
& m\,s$^{-1}$ \\

$\overline{w}$
& Mean lateral velocity
& m\,s$^{-1}$ \\

$w'$
& Instantaneous lateral velocity fluctuation
& m\,s$^{-1}$ \\

$w_{\mathrm{rms}}$
& Root-mean-square lateral velocity fluctuation
& m\,s$^{-1}$ \\

${w'}^2$
& Instantaneous squared lateral velocity fluctuation
& m$^2$\,s$^{-2}$ \\

$x_i$
& Cartesian coordinate in the $i$-th direction
& m \\

$x_j$
& Cartesian coordinate in the $j$-th direction
& m \\

$y$
& Vertical coordinate or distance from the bed
& m \\

$z_o$
& Roughness height
& m \\

$\Delta$
& LES filter width
& m \\

$\Delta t$
& Time interval / time step
& s \\

$\Delta x$
& Grid spacing in the $x$-direction
& m \\

$\delta_{\max}$
& Strength of velocity inversion
& \% \\

$\epsilon$
& Eddy viscosity associated with vertical momentum transport
& m$^2$\,s$^{-1}$ \\

$\nu$
& Molecular or kinematic viscosity
& m$^2$\,s$^{-1}$ \\

$\nu_t$
& Eddy viscosity introduced by the subgrid-scale model
& m$^2$\,s$^{-1}$ \\

$\rho$
& Fluid density
& kg\,m$^{-3}$ \\

$\sigma^2$
& Variance of the detrended velocity signal
& m$^2$\,s$^{-2}$ \\

$\tau_{\mathrm{bed}}$
& Bed shear stress
& Pa \\

$\tau_{ij}$
& Subgrid-scale stress tensor
& Pa \\

$\tau_{uw}$
& Reynolds shear stress component associated with $\overline{u'w'}$
& Pa \\

\end{longtable}
\section*{REFERENCES}
\printbibliography[heading=none]

\end{document}